\documentclass[twocolumn,showpacs,preprintnumbers,amsmath,amssymb,aps]{revtex4}

\usepackage{graphicx}
\usepackage{dcolumn}
\usepackage{bm}
\usepackage{textcomp}
\usepackage{pifont}
\usepackage{gensymb}
\usepackage{amsmath}

\begin{document}


\title{The Knight field and the nuclear dipole-dipole field in an InGaAs/GaAs quantum dot ensemble}

\author{T.~Auer$^{\star}$}
\author{R.~Oulton$^{\dag}$}
\author{A.~Bauschulte}
\author{D.~R.~Yakovlev$^{\ddag}$}
\author{M.~Bayer}
\affiliation{Experimentelle Physik 2,
             Technische Universit\"at Dortmund,
             44221 Dortmund, Germany}

\author{S.~Yu.~Verbin}
\author{R.~V.~Cherbunin}
\affiliation{Institute of Physics, St.~Petersburg State University, St.~Petersburg 198504, Russia}

\author{D.~Reuter}
\author{A.~D.~Wieck}
\affiliation{Angewandte Festk\"orperphysik,
Ruhr-Universit\"at Bochum, 44780 Bochum, Germany}

\date{\today}

\begin{abstract}
We present a comprehensive investigation of the electron-nuclear
system of negatively charged InGaAs/GaAs self-assembled quantum dots
under the influence of weak external magnetic fields (up to 2~mT).
We demonstrate that, in contrast to conventional semiconductor
systems, these small fields have a profound influence on the
electron spin dynamics, via the hyperfine interaction.  Quantum
dots, with their comparatively limited number of nuclei, present
electron-nuclear behavior that is unique to low-dimensional systems.
We show that the conventional Hanle effect used to measure electron
spin relaxation times, for example, cannot be used in these systems
when the spin lifetimes are
long. An individual nucleus in the QD is subject to milli-Tesla
effective fields, arising from the interaction with its
nearest-neighbors and with the electronic Knight field. The
alignment of each nucleus is influenced by application of external
fields of the same magnitude. A polarized nuclear system, which may
have an effective field strength of several Tesla, may easily be
influenced by these milli-Tesla fields. This in turn has a dramatic
effect on the electron spin dynamics, and we use this technique to
gain a measure of both the dipole-dipole field and the maximum
Knight field in our system, thus allowing us to estimate the maximum
Overhauser field that may be generated at zero external magnetic
field.  We also show that one may fine-tune the angle which the
Overhauser field makes with the optical axis.
\end{abstract}

\maketitle

\section{Introduction}

The expectation that the electron spin in semiconductor quantum dots
(QDs) could serve as a building block for quantum computing
applications has drawn renewed attention to the role of the quantum
dot nuclei. In the seventies it was shown that for bulk
semiconductors, the interplay between an electron spin (at that
time, of a donor trapped electron) and the nuclear spins in its
vicinity leads to a wide variety of effects and often exhibits
unexpected behavior
\cite{lampel1968,ekimov1972,ekimovsafarov1972,berkovits1973,dyakonovperel1974,dyakonov1974,fleisher1975,paget1977,OptO}.
In QDs the hyperfine coupling between electron and nuclear spins is
further enhanced by the strong localization of the electron in the
dot, giving rise to complex dynamics \cite{merkulov2009}.

Nuclear spins can be polarized by transfer of angular momentum from
optically oriented electrons, in a process known as the Overhauser
effect \cite{overhauser1953,lampel1968,berkovits1978}. It was shown
that nuclear polarization obtained in this way leads to an effective
magnetic field of the order Tesla for the QD electron
\cite{brown1996,gammon2001,koppens2005,akimov2006,eble2006,braun2006,tartakovskii2007,feng2007,maletinsky2007,oulton2007}.
Most of these experiments exploited the Overhauser energy shift
\cite{overhauser1953,brown1996} of the electron Zeeman levels split
in an external field where the nuclear field was of the same order
of magnitude as the external field.

In this paper we report experimental studies of InGaAs/GaAs quantum
dots.  We monitor the polarization of the QD ground state
photoluminescence (PL) in the presence of weak external magnetic
fields.  The magnitude of these fields (a few mT or less) are shown
to be far too small to have any direct effect on the dynamics of the
electron in the QD itself.  Milli-Tesla fields exerted onto electron
spins only affect the spin precession dynamics over timescales
greater than tens of nanoseconds.  It is only recently that electron
spin coherence times longer than this (several microseconds) have
been observed in QDs \cite{greilich2006}.  One might imagine that
milli-Tesla magnetic fields may be used in this long coherence time
regime to probe and manipulate spin dynamics, and in fact this has
been reported \cite{epstein2001,bracker2005}. However, as we will
demonstrate, considerable caution needs to be exercised in the
interpretation of such data, as interactions are usually present
which screen the direct effect of this external field on the
electron.

For the above reasons, the influence of milli-Tesla fields onto the
electron is usually negligible in semiconductors.  It may come as a
surprise, therefore, that we observe very dramatic effects on the
electron polarization when applying these fields.  This occurs in
our system due to the fact that the electron dynamics are governed
by the magnitude and direction of an effective nuclear, or
Overhauser field that is exerted onto the electron from $\sim$10$^5$
nuclei in the QD.  By polarizing a significant fraction of these
$\sim$10$^5$ nuclear spins in the same direction, one may generate
Tesla-strength Overhauser fields that completely dominate the
electron spin dynamics in the system.  The Tesla strength of these
Overhauser fields is however deceptive: the interaction is not a
real magnetic field, but an effective field that acts on the
electron only.

While the sum of the interaction from all of the nuclear spins onto
the electron is large, each nucleus itself is subject only to very
small effective fields. Considered from the point of view of a
single nucleus in the QD, the nucleus experiences three effective
fields (i) from its nearest neighbors (the dipole-dipole field), of
the order of magnitude of $\sim 0.1$~mT; (ii) from the electron, the
Knight field, of the order of magnitude 0.1 -- few mT; and (iii) any
external fields applied.  Thus in the regime of milli-Tesla applied
fields, it is easy to see that we begin to explore the competition
between the nearest neighbor field, the Knight field and the
external field.  We will demonstrate that by changing the magnitude
and direction of external milli-Tesla magnetic fields, one may
change the orientation of the Overhauser field.  By first optically
orienting the nuclear spins along the z-axis, and then applying a
milli-Tesla magnetic field, one may induce a change in orientation
of the entire Tesla-strength Overhauser field due to the precession
of each nuclear spin about the external magnetic field.

The manuscript is divided into five sections.  In Section II the
samples and experimental technique are outlined.  In Section III we
describe how angular momentum is transferred from circularly
polarized light to the electron-nuclear spin system in the QDs.  In
Section IV we consider theoretically the interaction between the
electron and the nuclei in this system.  Then in Section V the
experimental results are discussed.

Section V.A explores what happens when a magnetic field is applied
perpendicular to the optical axis.  It is found that in this
geometry, the applied field continuously reorients the nuclear spins
away from the optical orientation direction, and thus build-up of a
nuclear field is suppressed.  This only occurs, however, when the
applied field is larger than the nearest-neighbor interactions
between the nuclei, and we use this technique therefore to measure
the magnitude of the nearest-neighbor (dipole-dipole) interactions
in our system.

In Section V.B we apply a field that is equal and opposite to the
Knight field from the electrons, allowing us to obtain a value for
this field.  We refine the method used in previous work
\cite{lai2006} by using a combination of fields perpendicular
(transverse, $x$-direction) and parallel (longitudinal,
$z$-direction) to the optical axis, allowing the Knight field
feature to become more visible.  In Section V.C we estimate
theoretically the maximum degree of nuclear polarization we are able
to obtain in our sample.  Here we use the two important values we
have measured, the Knight and dipole-dipole fields, the two
quantities which govern spin diffusion in the system after the
nuclear spins have been polarized.  The Knight field acts to hold
the polarization, whereas the dipole-dipole field allows spin
diffusion.  The ratio of these effective fields governs the nuclear
polarization theoretically obtainable in our sample.  We demonstrate
that in principle, a nuclear polarization of up to $>$ 98 $\%$ may
be generated.

\section{Samples and Experiment}

The sample studied is a 20 layer InGaAs/GaAs self assembled QD
ensemble with a sheet dot density of $10^{10}$ cm$^{-2}$. 20 nm
below each QD layer a Silicon $\delta$-doping layer is located with
a doping density about equal the dot density. Thus each QD is
permanently occupied with on average one ``resident electron'', as
confirmed by pump-probe Faraday rotation measurements
\cite{greilich2006}. The InAs/GaAs QD heterostructure was grown by
molecular beam epitaxy on a (100) GaAs substrate. After growth it
was thermally annealed for 30 seconds at 900\textcelsius. This leads
to interdiffusion of Ga ions into the QDs, which shifts the ground
state emission to 1.34 eV \cite{fafard1999}.

The measurements were performed at a temperature of $T = 2$~K with
the sample installed in an optical bath cryostat which was placed
between three orthogonal pairs of Helmholtz coils allowing
application of external magnetic fields of a few mT in all
directions. The coils were used to compensate parasitic magnetic
fields, of e.~g.~geomagnetic origin as well as to apply fields up to
3 mT parallel or perpendicular to the optical axis (longitudinal,
$z$ direction or transverse, $x$ direction, respectively).

The optical excitation was performed using a mode locked Ti:Sapphire
laser with a pulse duration of 1.5 ps and a repetition rate of 75.6
MHz (pulses separated by 13.2 ns). The excitation energy was 1.459
eV which corresponds to the low energy flank of the wetting layer.
The helicity of the exciting light could be modulated by means of an
electro-optical modulator and a $\lambda/4$ wave plate.  With this
setup pulse trains with duration between 20 $\mu s$ and 500 ms of
$\sigma^+$ or $\sigma^-$ polarization were formed.

The beam was focused on the sample with a 10 cm focal length lens
which was simultaneously used to collect the PL. The
photoluminescence was dispersed with a 0.5 m monochromator and
detected circular polarization resolved with a silicon avalanche
photo-diode which was read out using a gated two-channel photon
counter. In order to ensure a homogeneous excitation of the QDs the
PL was collected from the center of the laser spot.

\section{Optical Orientation of electrons and negative circular polarization}

The QDs studied contain on average one ``resident'' electron. After
excitation of an electron-hole pair into the wetting layer and
subsequent capture of the carriers into the QD, a trion is formed in
these singly charged QDs. The trion ground state consists of two
electrons in the conduction band s-shell with antiparallel spins and
a single hole in the valence band s-shell. The helicity of the light
emitted after the trion decays is therefore governed by the spin
orientation of the hole.  As a consequence, the helicity of the
photon emitted directly determines the spin orientation of the
resident electron left in the QD after radiative recombination.

It is an established phenomenon for singly n-doped QDs under
non-resonant excitation and at zero magnetic field that the circular
polarization degree $\rho_{c}$ of the emission has the opposite
helicity to the excitation (known as negative circular polarization
effect, NCP). Here we use the standard definition:
\begin{equation}
\label{rhoc}
\rho_{c} = (I^{++} - I^{+-})/(I^{++} + I^{+-}),
\end{equation}
with $I^{++}$  denoting the intensity of PL having the same helicity as the excitation ($\sigma^{+}$) and $I^{+-}$ the intensity of PL polarized oppositely to the excitation.
\begin{figure}
\centering
\includegraphics[width=\columnwidth]{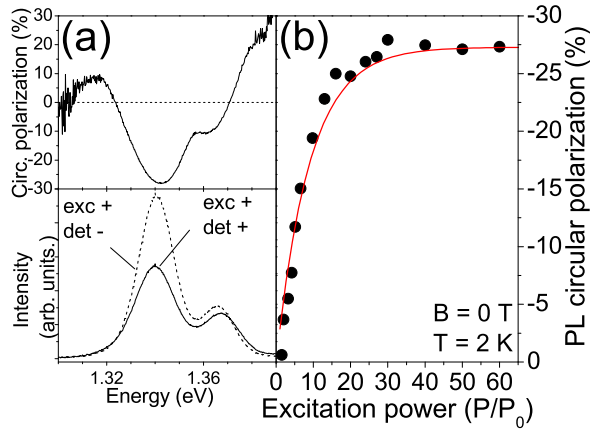}
\caption{(a) (color online) PL spectra of the QD sample studied (lower panel).
Excitation was $\sigma^+$ at 1.459 eV. The PL intensity of
$\sigma^-$ emission is greater than that of $\sigma^+$ emission,
thus the circular polarization degree is negative (upper panel). (b)
Power dependence of the circular polarization degree $\rho_c$.
$\rho_c$ rises with power and finally saturates indicating that a
memory of the spin orientation of the resident electrons is kept
until the following excitation cycle (see text). $P_0 = 2.5$
W/cm$^2$.} \label{figure1}
\end{figure}

Figure \ref{figure1} (a) depicts PL spectra of the QD ensemble
studied under $\sigma^+$ excitation in the wetting layer, with
detection either co-polarized ($\sigma^+$) or cross-polarized
($\sigma^-$) to the excitation. Both PL spectra show two peaks
corresponding to the inhomogeneously broadened ground state emission
at $\sim 1.34$ eV (s-shell), and to the first excited state emission
at $\sim 1.37$ eV (p-shell). The polarization of the PL is negative
throughout the emission from the s-shell (Fig.~\ref{figure1} (a),
upper panel).

Different mechanisms explaining the NCP effect have been suggested
\cite{cortez2002,kalevich2005,ware2005,verbin2005,akimov2005,laurent2006,masumoto2006}.
In Ref.~\cite{laurent2006} a mechanism was proposed whereby the
anisotropic exchange between an excited state electron and the hole
induces a spin exchange or ``flip-flop'' process, in order to
overcome Pauli blocking of the ground state electrons with parallel
spins.  In Ref.~\cite{ware2005}, it was suggested that dark excitons
are preferentially captured from the wetting layer by the QDs.  We
note, however, that all of these mechanisms imply that the electron
remaining after trion recombination accumulates spin polarization.
This is the important feature we exploit in our experiments. The
mechanism leading to this accumulation of electron spin polarization
may still be subject to discussions, but is not decisive for our
studies.

As in our previous work \cite{oulton2007}, we make use of this fact.
We excite the system with 75.6 MHz pulsed circularly polarized
excitation, allowing pumping of the electron spin population to
occur.  The electron spin polarization level reached in the sample
is governed by the competition between the optical pumping rate and
the decay dynamics of the electron spins.  During our measurements,
we always choose an excitation intensity well into the saturation
regime.  This ensures that when measuring the change in electron
spin polarization, the effects we observe are due to changes in the
decay rate of the electron spins, and not the optical pumping rate.

Figure~\ref{figure1}~(b) shows the PL polarization as a function of
excitation density.  As expected, $\rho_c$ is strongly excitation
power dependent.  This dependence reflects the efficiency of the
optical excitation of the dots, i.~e.~the average time between two
excitation events.  As the excitation power is increased, the
electron spins become more polarized and $\rho_c$ increases until it
reaches the saturation value of -27 to -30~\%.  The negative circular
polarization is limited by the fraction of loaded QDs in the
ensemble and the spin memory of the photoexcited electrons upon
relaxation. Neutral excitons and biexcitons may also be created in
the ensemble and spin preservation during relaxation from the
wetting layer may not be perfect.  Both these facts will reduce the
$\rho_c$ value. Therefore, even with full resident electron spin
polarization, the value of $\rho_c$ will not reach -100~\%.  In the
NCP effect, both the photoexcited electrons that retain their
polarization, and the polarized resident electron contribute to the
negative polarization. One may estimate the circular polarization
$\rho_c$ in a simple model: $\rho_c = Fx[P+S_z(I)]$, where $P$ is
the average polarization of the photoinjected electron spins,
independent of excitation intensity and $S_z(I)$ is the average
polarization of the resident electron spin along the z-axis, and is
dependent on excitation power.  $F$ is the fraction of QDs in the
sample that are singly charged and $x$ is the fraction of negatively
circularly polarized photons emitted when either the photoexcited or
the resident electron is polarized.  Depending on the position on
the sample, $F$ may be as large as 0.5 so that the electron spin
polarization in the singly charged dots may actually be considerably
larger, as the NCP measured. Thus we see that there is a linear
dependence between NCP and resident electron spin polarization where
the polarization of the photogenerated electron spins merely adds an
offset. This is not crucial in our case as we solely discuss {\it
changes} in NCP and thus electron spin polarization, respectively.

\section{Orientation of nuclear spins and the electron-nuclear spin system}
\label{electronnuclei}
\subsection{Electron spin precession in the Overhauser field}
In this section, we discuss in general terms the interaction between
an electron in a QD and its constituent nuclei at $B_{ext} = 0$.
Recent theoretical
\cite{khaetskii2000,merkulov2002,imamoglu2003,stepanenko2006,deng2006}
and experimental
\cite{dzhioev1999,braun2005,akimov2006,maletinsky2007,greilich2007}
work has demonstrated that this is the key electron spin relaxation
mechanism in QDs.  In addition, strong Overhauser fields in QDs have
been directly measured \cite{brown1996,tartakovskii2007} and
inferred \cite{oulton2007}.  We consider first of all the hyperfine
interaction in QDs, and consider the effect of an Overhauser field
onto the electron spin system under different optical orientation
regimes.  We then discuss the factors which influence the magnitude
of the Overhauser field, before discussing the experimental results
in milli-Tesla fields.

As an electron inside a QD is strongly localized, the interaction
between electron spin and a nuclear spin is enhanced in comparison
to bulk semiconductors. A single electron populating the conduction
band of a QD has a Bloch wavefunction with s-symmetry, leading to a
high electron density at the nuclear site.  The envelope
wavefunction has an overlap with about $10^5$ QD nuclei. This can be
estimated by considering an approximately disk shaped QD 20 nm in
diameter and 5 nm in height. Their spins interact with the electron
spin via the hyperfine interaction, described by the Fermi contact
Hamiltonian \cite{abragam1983}
\begin{equation}
\hat{H}_{hf} = \sum_i A_i |\psi (\mathbf{R}_i)|^2 \,\hat{\mathbf{S}} \cdot\hat{\mathbf{I}}_i.
\label{HF}
\end{equation}
Here $|\psi (\mathbf{R}_{i})|^2$ is the probability density of the
electron at the location $\mathbf{R}_{i}$ of the $i^{th}$ nucleus,
$A_{i}$ is the hyperfine interaction constant and
$\hat{\mathbf{S}}$, $\hat{\mathbf{I}}_{i}$ the operators of the
electron spin and the nuclear spin, respectively.

As we see from Eq.~(\ref{HF}), the total interaction energy is
dependent on the electron spin, $\mathbf{S}$, and the orientation of
each of the nuclear spins, $\mathbf{I}_i$.  The interaction energy
therefore crucially depends on the alignment of the nuclear spins in
the QD.  One may consider that the nuclei exert an effective
magnetic field $\mathbf{B}_N$ onto the electron as given by
\begin{equation}
\left( \sum_i A_i |\psi (\mathbf{R}_i)|^2 \mathbf{I}_i \right)\cdot \mathbf{S} = g_e \mu_B \mathbf{B}_N\!\cdot\!{\mathbf{S}}.
\label{BN2}
\end{equation}
where $g_e$ is the electron g-factor, $\mu_B$ the Bohr magneton.

\begin{figure}
\centering
\includegraphics[width=\columnwidth]{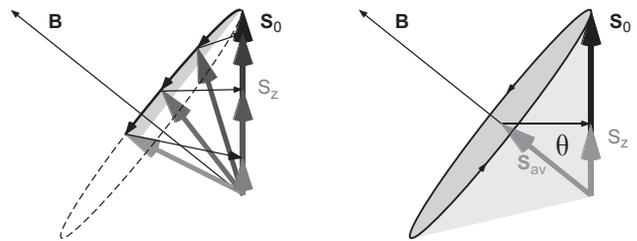}
\caption{Schematic of electron spin precession in a nuclear field
oriented at angle $\theta$ to the z-axis.  The average electron spin
vector is tilted at $\theta$.  The polarization measured, however is
given by the projection of this average electron spin onto the
$z$-direction, such that $S_z = S_0 \cos^2 \theta$.} \label{figure2}
\end{figure}

In light of the assumption that it is the hyperfine interaction that
causes electron spin decay, we neglect other spin dephasing
mechanisms (such as phonon interactions) at low temperatures and
consider what happens to an electron spin in this nuclear magnetic
field.  The excitation pulse induces formation of a trion, which
radiatively recombines to leave behind an electron spin polarized
along the $\pm z$ direction (the optical axis).  The electron spin
dynamics are then governed by the effective field $\mathbf{B}_N$. In
general, the motion of a spin $\mathbf{S}(t)$ in a fixed magnetic
field $\mathbf{B}$ is described by
\begin{eqnarray}
\label{precess}
\mathbf{S}(t) &=& (\mathbf{S}_0 \mathbf{\cdot b})\mathbf{b} + \left[ \mathbf{S}_0 - (\mathbf{S}_0 \mathbf{\cdot b})\mathbf{b} \right] \cos \omega t
+ \nonumber\\
&& \left[\left( \mathbf{S}_0 - (\mathbf{S}_0 \mathbf{\cdot b})\mathbf{b} \right) \times \mathbf{b} \right] \sin \omega t,
\end{eqnarray}
where $\mathbf{S}_0$ is the initial electron spin, $\mathbf{b} =
\mathbf{B}/B$ is the unit vector in direction of the magnetic field
acting on the electron and $\omega$ is the Larmor precession
frequency given by $\omega = |g_e| \mu_B B/\hbar$.  The dynamics of
the electron spin at $B_{ext} = 0$ is clearly dependent, therefore,
on the direction of the Overhauser field, which may have any
orientation in space.  If the Overhauser field is aligned along the
optical ($z$) axis, the electron spin will be static, and no
polarization will be lost.  For any $\mathbf{B}_N$ field not aligned
along $z$, however, the electron spin evolves in time.  In the
classical analogue, the electron spin precesses about the
$\mathbf{B}_N$ field.

In the measurements presented here, the electron spin precession
period in $B \approx 25$~mT (the order of magnitude of the frozen
nuclear spin fluctuation field, discussed below) is $\approx
5.7$~ns, compared to the 13~ns repetition period of the laser.  The
electron spin precesses more than twice before it is reinitialized.
{\bf Here, we assume that the QD in which the electron is confined is
excited by every pulse due to the high excitation intensity}. By the time it is sampled by the next excitation pulse in PL, we measure a time average of this electron spin in the ensemble. If we average the $z$ component of
$\mathbf{S}(t)$ over time we obtain
\begin{equation}
S_{z}(\mathbf{B})  =  S_0 \frac{B_{z}^2}{B^2} \quad =  S_0 \cos^2
(\theta(\mathbf{B})). \label{theta}
\end{equation}
The expression for ${S}_{z}$ is obviously analogous to the z
component of the projection of the initial electron spin $S_0$ on
the precession axis defined by the magnetic field where $\theta$ is
the angle between the precession axis and the $z$ axis (see
Fig.~\ref{figure2}).

Let us now consider what occurs in our system. Figure~\ref{figure2}
shows the precession dynamics of the electron spin. The circularly
polarized excitation results in spin population, giving an average
initial electron spin $S(t=0) = S_{0}$.  The electron precesses in
the Overhauser field several times and the ensemble reaches a steady
state of polarization $S_0 \cos^2 \theta$ before the next pulse
arrives 13.2 ns later.  This next pulse reads out the average
projection of the electron onto the $z$-axis, as discussed in
Section III.  Thus the PL polarization is dependent on the initial
electron spin $S_0$ and $\theta (\mathbf{B})$.  $S_0$ is governed by
the electron spin retained during energy relaxation from the wetting
layer to the QD ground state.  This is constant in our experiments,
as the excitation conditions are kept the same.  The angle of the
Overhauser field to the $z$-axis, $\theta(\mathbf{B})$, is the
factor that changes dramatically in these measurements, and we
consider now what governs the magnitude and orientation of the
Overhauser field.

\subsection{The nuclear magnetic field in the absence of optical orientation}
In this Section we discuss the electron spin dynamics for the case
where the nuclear spins are given no particular orientation.  One
might expect that in a system of randomly oriented nuclear spins,
the nuclear magnetic field would be zero, and the electrons thus
would be unaffected by the presence of the nuclei.  Generally,
however, the magnetic field generated by the sum of the nuclear
spins is never exactly zero.  The QD contains a large but
nevertheless finite number of nuclei ($N = 10^5$), which means that
statistically, the number of spins parallel and antiparallel in any
given direction will not be equal, but differ by a value
$\sqrt{N/3}$ at a particular ``snapshot'' in time.  The result is an
effective magnetic field $B_{f}$, orientated in a random direction
in 3D space, about which the electrons precess. The magnitude of
$B_{f}$ can be estimated by $B_{f} = b_{N}/\sqrt{N}$ with $b_{N}$
being the maximum nuclear magnetic field for 100 \% nuclear
polarization. We estimate below a value of $b_{N} = 8.3$ T for our
QDs. For $N \approx 10^5$ one thus obtains $B_{f} \approx 26$ mT
with an in-plane component of $B_{f,xy} \approx 20$~mT. Experimental
values of 10 to 30 mT for $B_{f}$  agree well with this estimate
\cite{ignatiev2007}.

How this $B_f$ field affects electron spin dephasing depends
crucially on the timescale of reorientation of the nuclear spins
compared to the precession period of the electron in $B_f$ (5 -- 6
ns for $B_f \approx 25$~mT).  Nuclear spin dynamics tend to be much
slower than electron spin dynamics: the nuclear spin fluctuation
field changes on a timescale of $10^{-6}$~s
\cite{braun2005,comment1} due to the precession of the nuclear spins
in the milli-Tesla magnetic field generated by the electron spin
\cite{merkulov2002} (see Table~\ref{times} for an overview about the
relevant timescales).  This means that over timescales less than
1~$\mu $s, the electron is exposed to a "snapshot" of $B_{f}$, where
the nuclear spin configuration remains ``frozen''.  In the absence
of an external magnetic field, only the internal field $\mathbf{B} =
\mathbf{B}_{f}$ acts on the electron.

The direction and the magnitude of this frozen nuclear spin
fluctuation will vary from dot to dot which leads to a rapid decay
of the average electron spin orientation in the ensemble (note that
this is also true for single dots when the electron polarization is
measured as an average over many excitation cycles exceeding the
nuclear fluctuation time).  Despite the fact that the $B_f$ field is
randomly oriented at any given time, the average electron spin
measured over the ensemble does not decay to zero.  Assuming that
the nuclear spins are randomly distributed, $B_{f,x} = B_{f,y} =
B_{f,z}$ and thus from Eq.~(\ref{theta}):
\begin{eqnarray}
\theta & = & \arccos{\frac{1}{\sqrt{3}}} \quad\widehat=\quad 54.7\degree, \nonumber\\
S_z & = & \frac{S_0}{3}.
\label{onethird}
\end{eqnarray}
For a randomly oriented nuclear spin system, the average angle of
$B_f$ is $\theta = 54.7\degree$, and the electron spin polarization
hence quickly decays to about 1/3 of its initial value due to the
frozen nuclear field. Total decay then follows on a microsecond
timescale due to continuous change in direction of this nuclear
field \cite{merkulov2002,braun2005}.  The value of 1/3 obviously
arises from the fact that the projection onto all directions in 3D
space is equal.  The initial orientation of the electron is
nevertheless important.  In this system the electron starts with
orientation along the $z$-axis but with no preferential direction in
the $x$-$y$ plane, so that, when ensemble averaging, a residual
projection onto the $z$-axis is retained, but no preferential
direction exists in the $x$-$y$ plane.

\subsection{Optical orientation of the nuclear spins}
\begin{table}
\caption{\label{times} Typical timescales occuring in the
electron-nuclear spin system assuming $g_e = 0.5$ {\bf at zero
external field}\cite{merkulov2002,OptO,paget1977}.}
\begin{ruledtabular}
\begin{tabular}{ll}
precession of electron spin in $\sim 10$~mT $B_f$ field& $\sim 10^{-9}$~s \\
precession of nuclear spins in Knight field & $\sim 10^{-6}$~s  \\
relaxation of nuclear spins in dipole-dipole field & $\sim 10^{-4}$~s \\
polarization of nuclear spins using NCP& $\sim 10^{-1}$~s\\
\end{tabular}
\end{ruledtabular}
\end{table}

Strong optical pumping of the system with circularly polarized light
leads to a continuous transfer of angular momentum from the photons
to the electron spins.  It is well-known that via spin flip-flops
with polarized electrons, orientation of the nuclear spins along the
axis of excitation may occur (Overhauser effect).  For this process
the respective orientation of the $B_f$ field in a QD at the point
of time of the trion decay may be important. For QDs containing a
$B_f$ field predominantly oriented along $z$ the electron spin's $z$
projection stays large and thus enough time is given to flip a
nuclear spin. However, in QDs where the $B_f$ field is by chance
predominantly transverse, the electron spin precesses and is not
able to polarize nuclear spins. While some $B_f$ configurations may
inhibit electron spin preservation, the nuclear spin system changes
on a microsecond timescale, so eventually most QDs will experience
some nuclear polarization.  Continuous optical pumping realigns the
electron after angular momentum transfer to a nucleus, such that
many nuclear spins become oriented. Without optical orientation, the
nuclear fluctuation fields in every QD of the ensemble are evenly
distributed in all three dimensions.  With optical orientation, an
additional field, the Overhauser field ${\bf B}_N$, is generated
along the $z$ axis, that may be much larger than the in-plane
component $B_{f,xy}$:
\begin{equation}
B_{N}  > B_{f,xy}.
\end{equation}
The electrons now precess about a nuclear field whose $z$ component
dominates, resulting in an increase of average electron spin
polarization $S_{z}$ in comparison to the case of a totally randomly
oriented nuclear system. The angle $\theta$ in Eq.~(\ref{theta})
decreases and $S_{z}$ increases.

A significant nuclear polarization obtained by strong pumping and
sufficiently long illumination of the sample leads to a nuclear
field $B_{N} \gg B_{f,xy}$ parallel to the $z$ axis and a marked
reduction of the influence of the nuclear fluctuation field,
maximally restoring electron spin alignment in $z$ direction
\cite{ignatiev2007}. Note that at the very edges of the QDs nuclear
spin diffusion out of the dot may occur, depending on the nuclear
species. In the core of the QDs, however, where the Knight field is
strong, spin diffusion is suppressed, and it is here that the
nuclear spins may become polarized in our case.

\begin{figure}
\centering
\includegraphics[width=\columnwidth]{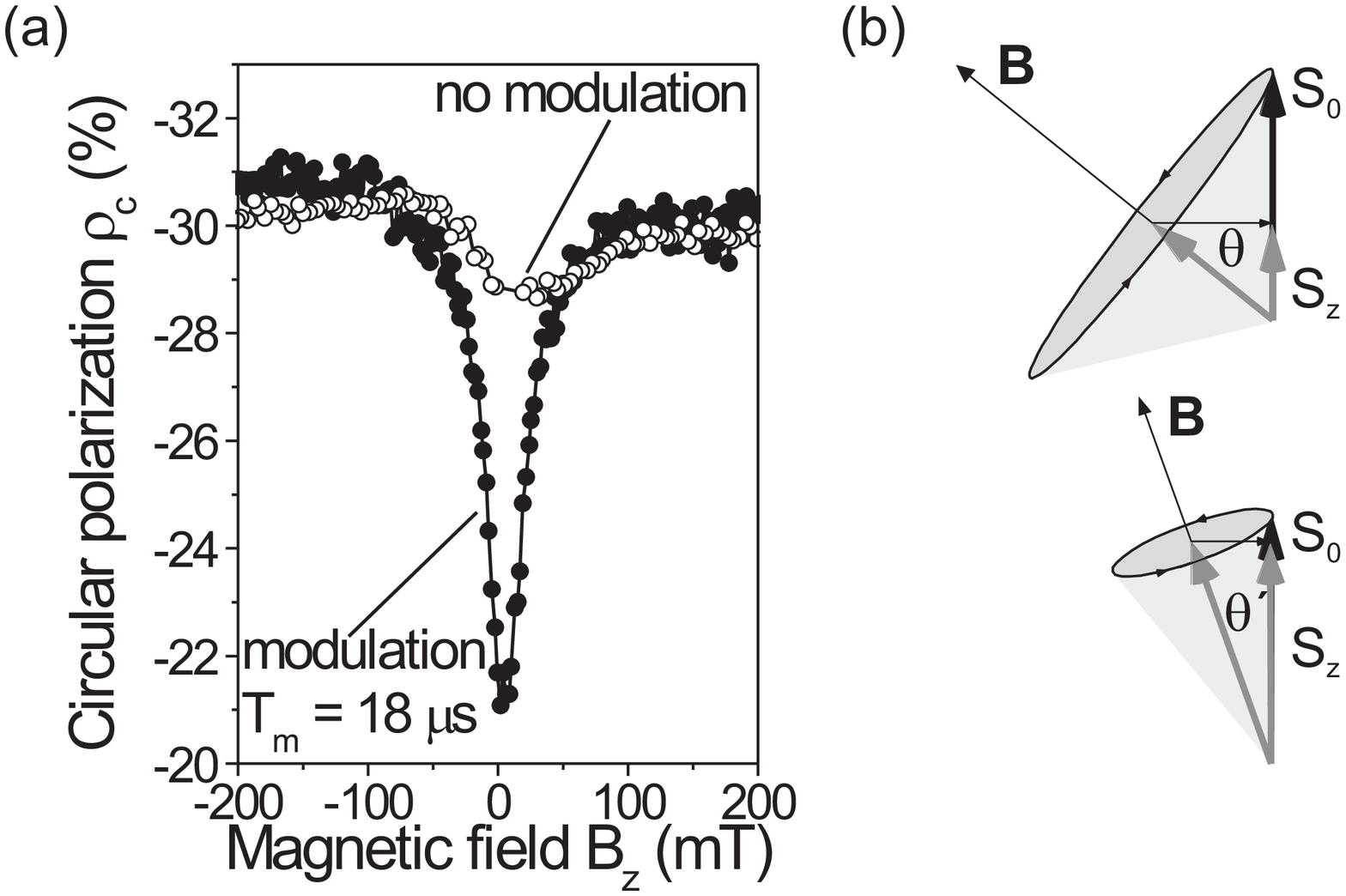}
\caption{(a) Dependence of the negative circular PL polarization on
a longitudinal magnetic field. For fast modulation of the excitation
polarization (full circles) no significant nuclear polarization can
build up and at $B_{z} = 0$ the nuclear fluctuation field lead to
depolarization of the ensemble average $S_{z}$. When $B_{z}$ becomes
large enough to suppress the nuclear fluctuation field $B_{f}$ the
polarization is maintained. Without modulation (open circles)
nuclear polarization in $z$ direction builds up and the resultant
$B_N$ field plays the same role as the external field, also at
$B_{z} = 0$. (b) Schematic for precession of the electron spin in
the fluctuation field: (top) as in Fig.~\ref{figure2}; and (bottom)
for a total field at a shallower angle to the $z$-axis. For
precession about the axis with $\theta^\prime < \theta$ more of
$S_z$ is conserved.} \label{figure3}
\end{figure}

Figure~\ref{figure3} (a) illustrates the effect of either an
externally applied field or an internal field generated by nuclear
polarization. Two curves are shown displaying the dependence of the
PL polarization on an external field in $z$ direction. One of them
was obtained for excitation helicity modulation between $\sigma^+$
and $\sigma^-$ with period $T_m = 18 \;\mu s$ which is about two orders
of magnitude faster than the nuclear spins need to be polarized
(with measurements performed during the $\sigma^+$ cycle only).
The other one was recorded with unmodulated excitation, i.~e.~constant
$\sigma^+$ excitation.  Let us first consider the modulated excitation case.
Here, the net angular momentum flux into the system averaged over time
is zero and no significant nuclear polarization $B_{N}$ builds up
(we will demonstrate in Sec.~V.A that significant nuclear polarization
requires tens of milliseconds pumping time). In this case, the $B_f$ field reduces the electron polarization at external field $B_{z} = 0$ to about $-21$ \%.
When $B_{z}$ is increased, the resultant field $\mathbf{B} = \mathbf{B}_{z} + \mathbf{B}_{f}$ is at an angle closer to the $z$ axis ($\theta^\prime < \theta$).
The projection of the electron polarization, given by Eq.~(\ref{theta}), increases, and $\rho_c$ goes from $\approx -21$ \% to $\approx -30$ \%.
For values $B_{z} \gg B_{f}$, $\theta^\prime \approx 0$ and the PL polarization saturates at $\approx -31$ \%. \\

We now compare this to the case where a nuclear field is allowed to
accumulate. Unmodulated excitation causes optical orientation of the
nuclear spins even at $B_{z} = 0$ and a nuclear field $B_{N}$ in $z$
direction builds up. This nuclear field plays exactly the same role
as an external field in increasing the projection of the electron
spin onto the $z$ axis. The resultant field onto the electron, given
by $\mathbf{B}_{f}  + \mathbf{B}_{N}$, is, again, closer to the $z$
axis. In this case, $B_{N}$ dominates over $B_{f}$. The polarization
reaches $\approx -29$~\% already at $B_z = 0$, demonstrating that for
almost all QDs, a significant nuclear polarization must occur.  Note
that the small dip which is still apparent for unmodulated
excitation in Fig.~\ref{figure3} is caused by the ensemble
distribution and the distribution of the $B_f$ field.  While a large
fraction of the QDs contains a polarized nuclear spin system there
may still be some with less nuclear polarization. For higher
external magnetic field,  almost all of the of QDs house a strongly
polarized nuclear spin system.

Note that in this work, the relative strength of the underlying
fluctuation field, $B_f$, and the optically generated Overhauser
field, $B_N$, hold the key to the electron spin dynamics.  In fact,
it is the presence of the significant fluctuation field as well as
an optically generated Overhauser field that are unique to QDs in
semiconductor systems.

\begin{figure}
\centering
\includegraphics[width=\columnwidth]{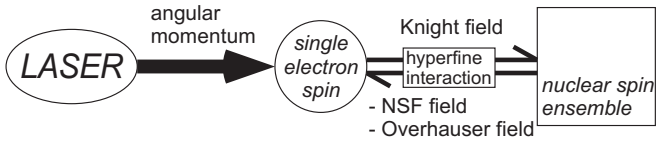}
\caption{Summary of the interactions in the electron-nuclear spin
system: A net angular momentum flux into the system is provided by
the circularly polarized laser light. On a single spin level the
electron and the nuclei interact via the hyperfine interaction. The
action of the entirety of the nuclear spins upon the electron spin
can be described by the Overhauser and the $B_f$ field. The nuclear
spins of the ensemble relax due to the dipole-dipole interaction
between them.} \label{figure4}
\end{figure}

Figure~\ref{figure4} summarizes all the different influences between
the resident electron spin, a single nuclear spin and the nuclear
spin ensemble.

\subsection{The influence of Knight field and dipole-dipole interaction on nuclear spin polarization}

We have discussed the effect of the nuclei on the electron spin
until this point by considering them en masse.  This approach is
sufficient for explaining the optical orientation of the nuclear
system as a whole, and the subsequent effect on the electron spin.
The electron-nuclear system is, however, a highly interdependent
coupled system, and to understand the electron spin dynamics one
must also gain a detailed understanding of the nuclear spin
dynamics.  In this section, therefore, we consider what happens to a
single nuclear spin as it interacts with its nearest neighbors, the
electron and an external field.

The nuclear spin system is relatively isolated, such that in the
absence of external magnetic fields, only two interactions dominate.
First is the nearest-neighbor interaction.  In an unpolarized
nuclear ensemble, each nucleus experiences an effective magnetic
field from the neighboring nuclear spins.  This field (also known as
the dipole-dipole field), denoted as $\tilde{B}_{L}$, is of the
order of $0.1$~mT (as we experimentally confirm later) and
fluctuates on the timescale of $10^{-4}$ seconds \cite{OptO}.  As we
will see in the experimental results in Section V.A., the timescale
for dynamic nuclear polarization is $10^{-2}$ seconds, two orders of
magnitude slower than the dipole-dipole interaction.  It has
nevertheless been observed that even in zero applied external field
conditions, significant nuclear polarization may occur in QDs
\cite{oulton2006,lai2006}.  The dipole-dipole interaction is not
spin conserving, which leads to the question of why nuclear
polarization may occur at all, and brings us onto the second
interaction.

The second important interaction that a nuclear spin has with its
environment is that with the electron (the hyperfine interaction).
Equation~\ref{rhoc} may be expanded to be expressed as:
\begin{equation}
\hat{H}_{hf} = \sum_i A_i |\psi (\mathbf{R}_i)|^2 \,[\hat{\mathbf{S}_z} \cdot\hat{\mathbf{I}}^i_z + (\hat{\mathbf{S}_+} \cdot\hat{\mathbf{I}}^i_- + \hat{\mathbf{S}_-} \cdot\hat{\mathbf{I}}^i_+)/2].
\label{HFexpanded1}
\end{equation}
The second term expresses the electron-nuclear spin flip-flop
interaction responsible for the optical orientation of the nuclear
spins, as described before.  The first term, on the other hand, may
be re-expressed as an effective magnetic field from the electron,
$\mathbf{B}^i_e$, acting onto the $i^{th}$ nuclear spin:

\begin{equation}
\hat{H}_{hf} = -\hbar \mu_i \sum_i \hat{\mathbf{B}^i_e} \cdot\hat{\mathbf{I}}^i_z + \sum_i A_i |\psi (\mathbf{R}_i)|^2(\hat{\mathbf{S}_+} \cdot\hat{\mathbf{I}}^i_- + \hat{\mathbf{S}_-} \cdot\hat{\mathbf{I}}^i_+)/2,
\label{HFexpanded2}
\end{equation}
where
\begin{equation}
\mathbf{B}^i_e = A_i |\psi (\mathbf{R}_i)|^2 \,\hat{\mathbf{S}_z}.
\label{defKnight1}
\end{equation}
This effective field, the ``Knight field'', was first identified in
nuclear magnetic resonance (NMR) experiments \cite{knight1949} as a
shift in frequency of the characteristic NMR resonance for
particular metals.  This shift occurs due to a paramagnetic effect
from the presence of conduction band electrons, with the magnitude
of this shift in energy being equal to the hyperfine splitting of
the ground state of the atom.  In our experiments, we cannot measure
the Knight shift directly without NMR techniques, and so, as we will
see in Section V, we apply a magnetic field equal and opposite to
this Knight field and investigate the back-action effect on the
Overhauser field.

From Eq.~\ref{defKnight1} we see that each nuclear spin experiences
a Knight field that depends on (i) the
location of the nucleus in the QD (the Knight field will be
strongest for a nucleus in the centre of the QD, where $|\psi
(\mathbf{R}_i)|^2$ is largest) and (ii) $S_z$, the projection of the
electron spin onto the $z$-axis.  We have already discussed in
Section IV.A that the electron precesses on a timescale of less than
10~ns, whereas the interaction of each nucleus is much slower than
this.  We may therefore use the time-averaged electron spin
projection $S_z = S_0 \cos^2\theta$, as given in Eq.~\ref{theta}.

In the absence of an external field, the value of the Knight field
is an important quantity in determining whether nuclear polarization
occurs.  The effective field generated by the electron acts to
screen the dipole-dipole interaction, and inhibits nuclear spin
diffusion \cite{OptO,lai2006}.  If one makes the assumption that the
maximum nuclear polarization may be achieved as long as
dipole-dipole diffusion is completely suppressed, one may express
the competition between the Knight field and the dipole-dipole field
as \cite{OptO,ThThesis}:
\begin{eqnarray}
\frac{B_N}{b_N} & \approx &  \frac{B_{e}^2}{B_{e}^2 + \tilde{B}_L^2}. \label{knight_simple}
\end{eqnarray}
where $b_N$ is the maximum achievable nuclear field for a given alloy system.  Thus, the value $B_e$ is an important one: a Knight field value significantly larger than the dipole-dipole field will allow nuclear polarization to occur.  Due to Knight field variation, nuclear polarization will obviously vary across the QD.

Going back to Eq.~\ref{defKnight1} we see that the Knight field is also dependent on the electron spin polarization, $S_z$.  We will see later that the experimentally determined value of the Knight field is dependent on the electron spin polarization, but it is also useful to define the maximum Knight field, $b^i_e$ for a given nucleus, such that:
\begin{equation}
\mathbf{B}^i_e  = -b^i_e \frac{\langle \mathbf{S}\rangle}{S}. \label{knight1}
\end{equation}
where $b_e^i$ is the maximum Knight field at a particular nuclear
site \cite{paget1977,OptO}.  The maximum Knight field is, in fact,
the more important quantity to be determined in a particular QD system.
As long as a strong nuclear field is generated ($B_N \gg B_f$) an
electron will remain aligned along the $z$-axis, and the Knight
field at all the nuclei will be the maximum possible in that system.
As we will see later, in our experiments the only technique we have
to probe the Knight field is to apply a field equal and opposite to
it in order to depolarize the nuclei.  At this point the average
electron spin decreases significantly, and this must be taken into
account when determining the experimental value of the Knight field
value that we measure.  However, although we measure $B_e$, we may
extrapolate $b_e$ because we also have a value for $S_z$.

Finally, it is also important to note that in our experiment we
determine a "weighted average" value of the Knight field.  The
Knight field value will vary between each nucleus, going from a
maximum value in the centre ($b^{max}_{e}$) to zero outside the QD.
Making the approximation that $|\psi (r)|^2 = \exp(-2r/a_0)$, where
$a_0$ is the radius of the QD and $r$ is the distance from the QD
centre, one may estimate that the average value is equal to half the
maximum Knight field $b^{max}_{e}/2$ \cite{foot1}.  In Section V we
will show that we measure a value of the Knight field, $\tilde B_e$,
that is an average of the entire nuclear spin ensemble in the QD,
and is also dependent on $S_z$.

\section{Results and Discussion}

\subsection{Hanle measurements and the dipole-dipole field}
\label{sechanle}
In this Section we discuss the application of a
purely transverse magnetic field.  We discuss the often-used
interpretation of this type of experiment for determining the
electron spin relaxation time, and demonstrate experimentally that
for a QD system, this interpretation does not hold, and that in
general it cannot be used for Hanle curve widths $<$~10's~mT.  We
instead demonstrate that we may use this technique to determine the
strength of nuclear dipole-dipole interactions in the system.

The Hanle effect \cite{hanle1924} is a technique often used in
semiconductor physics to determine spin relaxation times by
initializing spin in the system in a particular direction
(e.~g.~along the $z$-axis) at time $t = 0$  and monitoring the
decrease of PL polarization as a transverse (known as Voigt
geometry) magnetic field is applied \cite{OptO}.  An electron spin
will precess in the $z$-$y$ plane, such that, according to
Eq.~(\ref{precess}), the dynamics will be given by $S_{z}(t) = S_0
\cos(\omega t)$.  When integrating over many cycles, the projection
of the spin onto the $z$-axis is clearly zero for integration times
considerably smaller than the spin lifetime.  For a finite electron
spin lifetime, $T_S$, however, the dynamics is given by $S_{z}(t) =
S_0 \exp(-t/T_S) \cos(\omega t)$.
\begin{figure}
\centering
\includegraphics[width=\columnwidth]{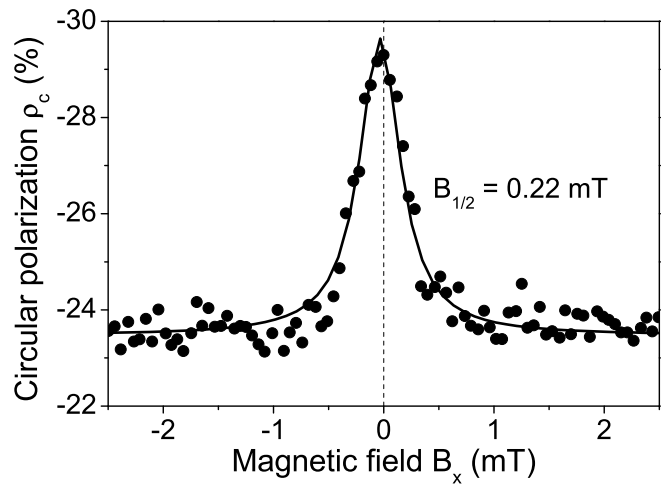}
\caption{Dependence of the PL circular polarization degree on a transverse magnetic field $B_{x}$ under fixed circularly polarized excitation (full circles: measured data, solid line: Lorentz fit). When the external field is strong enough to overcome $\tilde{B}_{L}$ the nuclear spins precess about the external field leading to a reorientation of the Overhauser field along the external field. The electron spins become depolarized by precession about the nuclear field. In practice, however, for $B_{x} > \tilde{B}_{L}$ nuclear polarization is no longer generated  and the electron spin is left with the fluctuation field (see text).}
\label{figure5}
\end{figure}

The Hanle effect \cite{hanle1924} may be used in a quantitative
manner to determine the spin lifetime.  The PL polarization is
monitored as a Voigt geometry magnetic field is increased.
Monitoring $S_{z}$ via the PL polarization whilst sweeping the
transverse field yields a Lorentzian curve, the width of which is
inversely proportional to the $T_S$ time of the electron:
\begin{equation}
\label{hanle}
T_S = \frac{\hbar}{|g_{e,x}| \mu_0 B_{1/2}},
\end{equation}
with the electron g factor $g_{e,x}$ along $x$, and the measured half width at half maximum $B_{1/2}$ of the Lorentz curve obtained.

Figure~\ref{figure5} shows a graph obtained from a Hanle measurement
under unmodulated excitation, with pulses of 75.6 MHz repetition
rate on our QD sample. A transverse magnetic field $B_{x}$ was swept
and the PL polarization for each field value was recorded. The PL
polarization drops sharply from its maximum value at $B_{x} = 0$ the
half-width of the peak being $\sim$ 0.22 mT.  From the usual
interpretation of the Hanle effect, Eq.~(\ref{hanle}), this would
correspond to $T_S \sim 57$~ns assuming $g_{e,x} = 0.5$
\cite{greilich2006}.

\begin{figure}
\centering
\includegraphics[width=\columnwidth]{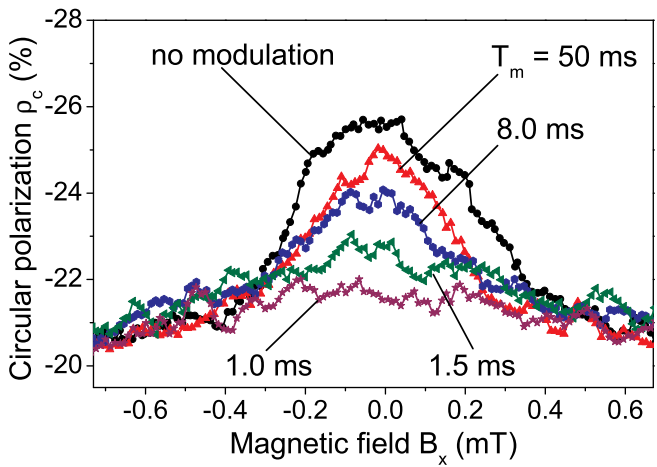}
\caption{(color online) (a) Hanle curves for excitation modulated between $\sigma^+$ and $\sigma^-$ with period $T_m$ (smoothed). The narrow peak gradually vanishes for faster modulation and is almost absent for $T_m = 1$ ms.}
\label{figure6}
\end{figure}

Let us now consider what happens if the excitation helicity is
flipped between $\sigma^+$ and $\sigma^-$ with period $T_m$  that we
are able to vary over a range of milliseconds. Fig.~\ref{figure6}
shows Hanle curves under excitation modulated in such way (note that
because the signal could only be measured during the $\sigma^+$
pulse, the measurement is inherently more noisy).  The narrow peak
appearing with unmodulated excitation at $B_x = 0$ gradually
disappears when the modulation frequency is increased.  It almost
completely vanishes for a modulation period of $T_m = 1$ ms.  This
value is in complete disagreement with the value obtained from
Eq.~(\ref{hanle}): if the electron spin relaxation time is of the
order of $T_2^{\star} \sim 57$~ns, the slower 1~ms modulation should
not inhibit it.  In general, as the dynamics of the electron spin
takes place on a nanosecond (precession) to microsecond (coherence)
timescale (see Table~\ref{times}), this millisecond effect is far
too long to be of electronic origin.  Although transverse fields
also lead to the depolarization of polarized electron spins in the
QDs under study, the situation obviously fundamentally differs from
the one underlying the original Hanle effect.

When considering the discussion in Section IV.B, it becomes
immediately clear that the Hanle effect as discussed above should
never be observed in this low field regime.  In fact, it appears at
first surprising that a dramatic change in the PL polarization is
observed at all.  As discussed in Section IV.B, the fluctuation
field from the nuclei, $B_f \sim 10$'s~mT, is always present and
should screen completely the effect of any milli-Tesla applied
field.  Polarization of the nuclear spins may occur, but this will
always result in an increase in the magnitude of the total nuclear
magnetic field, $|\mathbf{B}_f + \mathbf{B}_N|$, which will always
screen any mT applied field.  Thus, care has to be taken when the
width of the Hanle peak is used to determine the spin lifetime of
the electron in QDs as has been suggested in the literature
\cite{epstein2001,bracker2005}. This works only for nuclear
fluctuation fields $B_f \ll 10$~mT. At fields lower than several
tens of milli-Tesla the Overhauser field or the frozen fluctuation
field will always dominate the spin dynamics in QDs.

The fact that an effect is only observed when tens of milliseconds
excitation is used implies that the effect arises from dynamic
nuclear polarization which is known to occur on these timescales
\cite{gammon2001,maletinsky2007}.  In the following we will explain
the mechanisms leading to the specific shape of the curve in
Fig.~\ref{figure6}.

For modulated excitation with $T_m < 1$~ms, almost no nuclear
polarization is allowed to build up, and the electron feels the
fluctuation field only.  The electron spin decays to give $\rho_c
\sim -21\%$ irrespective of the applied field.  For unmodulated
excitation on the other hand, a nuclear magnetic field builds up
along the $z$ axis when $B_{x} = 0$. The nuclear field adds to the
frozen fluctuation field $B_{f,z}$ and reduces the angle $\theta$
between the $z$ axis and the total nuclear magnetic field
$\mathbf{B}_{N,tot} = \mathbf{B}_{N} + \mathbf{B}_{f}$ as discussed
in Sec.~\ref{electronnuclei}.  The PL polarization, $\rho_c \propto S_z$ increases from $\sim
-21$~\% to $\sim -26$~\% \cite{comment2}, similar to the behavior
shown in Fig.~\ref{figure3}.

Now let us consider what happens when a transverse field $B_{x}$ of
a few mT is switched on.  The electron is not sensitive to this
field as it is screened by the much stronger nuclear fields (the
fluctuation field or the Overhauser field), and so at first, it will
continue to polarize the nuclear spins in the $z$-direction.  The
nuclear spins, however, are sensitive to this transverse field.  As
discussed in Section IV.D, the nuclei feel three fields: the
external field $B_x$, the Knight field $B_e$ and the dipole-dipole
field $\tilde{B}_L$, all of which are of the same order of
magnitude.  If the external field dominates over the other two, then
over the timescale of $\mu$s, the nuclear spins begin to precess
about this field.

This situation has been investigated in detail in
Ref.~\cite{paget1977} for donors in bulk GaAs.  In this work, it was
discussed that application of a magnetic field in an oblique
direction results in the optically oriented nuclear spins precessing
about this external field effectively aligning the Overhauser field
along it.  The electron still experiences this Overhauser field
which is, however, now oriented along the external field: the
Overhauser field effectively magnifies the external field by several
orders of magnitude.  This was described in reference
\cite{paget1977,OptO} as:
\begin{equation}
\label{kappa}
\mathbf{B}_N = \kappa(\mathbf{S},\mathbf{B}_{ext}) \,\mathbf{B}_{ext},
\end{equation}
where $\kappa$ is known as the multiplication factor.  Note that the
polarization of the nuclear spins only occurs due to optical
orientation via the spin polarized resident electron, whereas the
external field solely directs the optically generated Overhauser
field. If an Overhauser field of several Tesla is generated and
realigned to the external milli-Tesla magnetic field, $\kappa$ can
reach values of $10^3$ or more.  This extraordinary effect is the
reason why such small external fields can have such a dramatic
effect on the electron spin system.

For a purely transverse field, the polarized nuclear spins will
precess about the $x$-direction in a few $\mu$s.  This has the
subsequent effect of destroying the electron spin polarization: it
will precess about the $x$-axis and all projection onto the $z$-axis
is lost.  At this point, no further nuclear polarization via the
electron spin can occur.  The nuclear polarization already present
will diffuse.  Thus we see that dynamic nuclear polarization cannot
occur in the steady state when a $B_x$ field is applied.

We observe though that a finite applied field is required to reduce
the $\rho_c$.  In Fig.~\ref{figure5} the PL value drops from $\sim$
-29$\%$ to $\sim$-24$\%$ steadily over the range of $\sim$0.6~mT and
then does not decrease further.  This fact may be explained by the
presence of the dipole-dipole field $\tilde{B}_{L}$.  For external
magnetic fields $B_x < \tilde{B}_L$ the external field is screened
by $\tilde{B}_L$. In order to realign the nuclear spins, $B_x$ has
to dominate over the dipole-dipole field.

For measurements in a purely transverse magnetic field the $x$
component of the Knight field is zero, thus the only transverse
field experienced by the nuclear spins is the external field. The
width of the Hanle peak is hence solely determined by the
competition between $\tilde{B}_{L}$ and $B_{ext}$. This is discussed
further in \cite{OptO, paget1977}. In this regime the width of the
depolarization peak is a measure for the dipole-dipole field
$\tilde{B}_{L}$ as discussed above. The average peak half width from
several measurements corresponds to a dipole-dipole field of
\begin{equation}
\tilde{B}_L = 0.22 \pm 0.02 \;\text{mT}.
\end{equation}

We shall now examine what happens when $B_x$ is increased above the
magnitude of $\tilde{B}_L$ and why the polarization remains at a
constant level for $B_x \gtrsim$ 0.6 mT, and does not drop to zero,
as may be expected if the field the electrons experience is purely
transverse.  To understand this behavior, one must consider the
magnitude of the nuclear polarization field $B_{N}$ relative to the
fluctuation field $B_{f}$.  The field $B_{N}$ will decrease as
$B_{x}$ increases: this is due to the fact that $B_{N}$ is dependent
on $S_z$ and decreases with decreasing electron spin because then
the ability of the electron spin to polarize the nuclear spins is
reduced. Thus, at a sufficiently large $B_{x}$, the Overhauser field
is close to zero, and only the fluctuation field $B_f$ remains.  The
fluctuation field then dominates the electron spin dynamics, and in
Fig.~\ref{figure5} the $\rho_c$ value reaches $\approx$ -23 \%, the
value found at 0 T in Fig.~\ref{figure3}~(a) for modulated
excitation (i.e. with no nuclear polarization).

\subsection{The Knight field}
We have seen from Sec.~V.A that the nuclei are sensitive to
extremely small fields.  In Section~\ref{electronnuclei} we also
discussed the importance of the Knight field in allowing nuclear
polarization.  The Knight field $\tilde{B}_e$ is the effective
magnetic field felt by each nucleus from the resident electron.
$\tilde{B}_e$ is antiparallel to the electron spin, and in our
scheme, is thus parallel to the $z$-axis.  At $B_{ext} = 0$ the
Knight field screens the effect of the dipole-dipole field and
allows dynamic nuclear polarization to occur; however,
$\tilde{B}_e$ must be stronger than $\tilde{B}_L$ \cite{lai2006,
paget1977}.

The magnitude of the Knight field is a quantity which varies not
only between different QDs, but also between individual nuclei in a
single QD, as its magnitude is proportional to the density of the
electron wavefunction at a particular nuclear site.  For QDs, the
Knight field may be an order of magnitude stronger than in bulk, due
to the increased electron density over fewer nuclei in the QD.  This
is why dynamic polarization in QDs at $B_{ext} = 0$ can be much
stronger than in bulk material.  It is therefore of great interest
to gain a measure of the strength of this effective field.

An approximate measure of the Knight field in QDs was determined for
the first time by Lai et al.~\cite{lai2006}.  In this measurement,
the PL polarization of a single QD exciton state was measured as a
milli-Tesla field was swept along the $z$ direction.  It was found
that a dip in the polarization was visible at $\sim 0.6$~mT, whose
position changed sign as the helicity of the excitation was changed.
This decrease in polarization is due to the fact that the external
field applied is exactly equal in magnitude and opposite in
direction to the Knight field in that point, i.~e.~$B_z =
-\tilde{B}_e$.  A nucleus in the QD then experiences an approximate
cancelation of the Knight field with the external field.  Without
the Knight field, dipole-dipole depolarization of the nuclear spins
occurs quickly, and dynamic nuclear polarization does not build up.

\begin{figure}
\centering
\includegraphics[width=0.6\columnwidth]{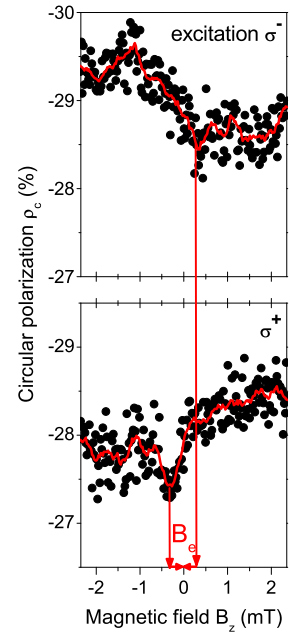}
\caption{(color online) $B_{z}$ dependence of the PL polarization ($\rho_c$) for $B_x = 0$ and fixed excitation polarization (full circles: measured data, solid line: smoothed). The dip at $B_{eff} = 0$ is shifted from $B_z = 0$ by the value of the Knight field $\tilde{B}_e$. Note that the sign of $\rho_c$ is opposite to the circular polarization of the excitation, see Eq.~\ref{rhoc}.}
\label{figure7}
\end{figure}

An identical measurement is performed in our system in the strong
pumping regime, with the results shown in Fig.~\ref{figure7}.  Here,
the $B_{z}$ dependence of the polarization is depicted for both
excitation helicities. The PL polarization exhibits a barely
discernible dip which is offset from $B_{z} = 0$ and whose position
is reversible with helicity.  As in Ref.~\cite{lai2006}, the shift
corresponds to the external field which is needed to compensate the
Knight field.  However, the effect is very small.  This is due to
the fact that in an ensemble the Knight field is fairly
inhomogeneous, and it is very difficult to depolarize all of the
nuclear spins at the same time.

In our previous work \cite{oulton2007} we presented evidence that we
achieve extremely large Overhauser fields ($> 7$~T for some QDs) in
our system.  If this strong field is aligned along the $z$ axis, the
polarization will be independent of the magnitude of $B_N$: as
discussed in Section~IV.C.  As long as $B_N \gg B_f$ and aligned
along the $z$-axis, the electrons do not depolarize.  In order to
see a visible effect when applying $B_z = -\tilde{B}_e$, one must
reduce $B_N$ to be of the same order of magnitude as $B_f$. For an
Overhauser field of a few Tesla, this means that $\sim 99 \%$ of the
nuclei contributing to this field must be depolarized
simultaneously. We therefore use a method to measure the Knight
field that was first reported in 1977 for electrons on donors
\cite{paget1977, OptO}, and present this method as ideal one for
investigating the Knight field in QDs.
\begin{figure}
\centering
\includegraphics[width=\columnwidth]{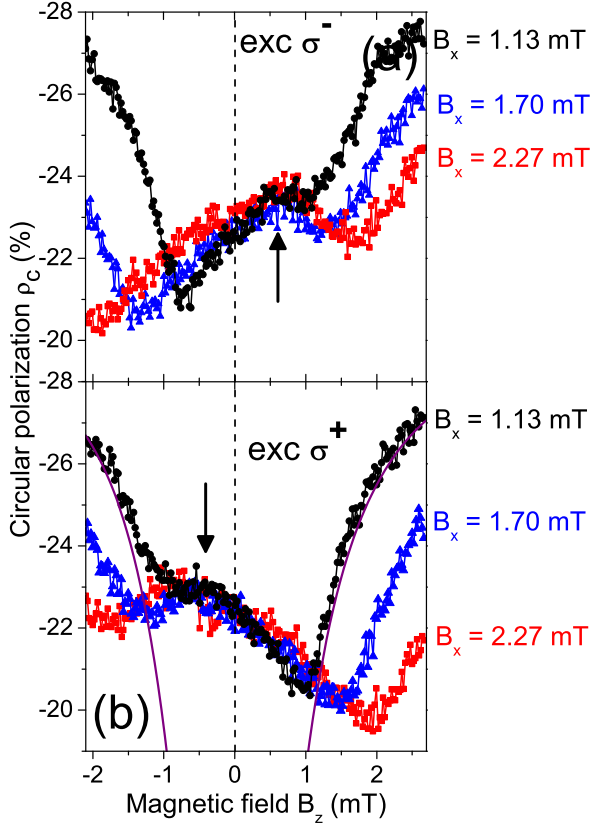}
\caption{(color online) $B_{z}$ dependence of the PL polarization for different values of $B_{x}$: 1.13 mT, 1.7 mT and 2.27 mT. Excitation $\sigma^-$ (a) and $\sigma^+$ (b). The constant shift of the peak in the middle of the W-shaped structure (indicated by arrows) corresponds to the magnitude of the Knight field. Solid line in (b): Polarization behavior expected when the Knight field and the fluctuation field are neglected.}
\label{figure8}
\end{figure}

In order to make the Knight field more visible, we additionally
apply a constant transverse magnetic field $B_{x}$, and again step
the $B_{z}$ field.  This technique, used extensively in
\cite{paget1977}, allows one to access nuclear polarization regimes
where $B_{N}$ is smaller, without significantly changing the
electron spin polarization generated (as would be the case, for
example, when decreasing the excitation power).  In this technique,
a transverse field is applied of magnitude 1.13 to 2.27 mT.  The
$z$-field is then swept, and the polarization is measured.  By
keeping the $B_{x}$ field constant and sweeping $B_{z}$, one
effectively sweeps the angle, $\theta_{ext}$ that the total field
$\mathbf{B}_{tot}=\mathbf{B}_{x}+\mathbf{B}_{z}$ makes with the
$z$-axis, given by $\theta_{ext} = \arctan(B_x/B_z)$.  In fact, we
will demonstrate that by using this technique, the angle that the
$\mathbf{B}_N$ field makes with the $z$-axis may be finely tuned,
and even become closer to the $x$-axis than $\theta = 54\degree$,
the angle at which the ensemble averaged frozen fluctuation field is
generally tilted from the $z$ axis.

Let us consider what should be expected as the $B_z$ field is swept
in the presence of a $B_x$ field.  By sweeping the $B_z$ field, the
angle of the total applied field is swept from $90\degree$ for $B_z
= 0$ to close to $0\degree$ when $B_z \gg B_x$.  As in the rest of
the work presented here, the applied field has no direct effect on
the electron, but each of the nuclei will respond to this field, and
in the absence of other effects, precess about the axis at an angle
$\theta$.

Now let us assume that strong dynamic nuclear polarization is
generated, where $B_N \gg B_f $.  As we have discussed previously,
the electron spin polarization is governed by this field: $S_z =
S_0\cos^2 \theta (B_N)$.  Thus sweeping the field should reveal a
change in circular polarization from 0 to $-100 \%$ if a strong $B_N$
field is created.  In Fig.~\ref{figure8}, $B_{z}$ dependencies for
various applied transverse fields are shown for both $\sigma^+$ and
$\sigma^-$ excitation.  We do not observe the change from 0 to
-100~\%, despite the fact that clear changes in PL polarization occur
on sweeping the field.  A clear asymmetry is present, however, that
is reversed upon reversal of the excitation polarization.

It is not surprising that the PL polarization does not drop to zero
for low $B_z$ values.  As in Fig.~\ref{figure5}, for $\mathbf{S}_0
\bot \mathbf{B}$, no nuclear polarization can occur. The electron
spin, however, is still exposed to the nuclear fluctuation field
which does not fully depolarize the electron spin.  The value
$\rho_c \sim $ -23 $\%$, is the same value as for $B_x \gg 0.22$~mT
in Fig.~\ref{figure5}, as we expect.

The solid line in Fig.~\ref{figure8} (b) shows the polarization
behavior expected when the Knight field and the fluctuation field
are neglected, and if we were to assume that $B_N$ is parallel to
$B_{ext}$.  The $B_{N}$ field direction would vary from $\theta =
90\degree$ at $B_{z}= 0$ T to $\theta = 0\degree$ at $B_{z} \gg
B_{x}$, and the time averaged electron spin $\langle \mathbf{S}
\rangle$ would follow it also.  In fact, as the $B_N$ field angle
moves away from the $z$ axis, its magnitude decreases and the
fluctuation field $B_f$ begins to dominate.  For this reason the
data do not follow the solid line.

The PL polarization exhibits a pronounced asymmetric W-shaped
behavior on sweeping the longitudinal field, that is inverted on
changing the excitation helicity.  Let us consider first the points
indicated by arrows in Fig.~\ref{figure8}, corresponding to local
turning points in the curve.  These occur at $\sim \pm 0.5$~mT.  We
note that at these points, $\rho_c$ reaches a value of $\approx -23$
\% for all the curves, and moreover, that these points,
approximately correspond in magnitude and sign to the Knight field
values observed in Fig.~\ref{figure7}.  At this point, the $B_{z}$
field approximately compensates the Knight field for many of the QD
nuclei. Due to the cancelation of the Knight field the nuclear spins
experience a purely transverse magnetic field. In this geometry
nuclear polarization is not allowed to built up as $\mathbf{B}_{ext}
\bot \mathbf{S}_0$. Thus the $\rho_c$ value measured corresponds to
the fluctuation field value of $\approx -23$~\%.

The marked asymmetry in the curves that are inverted when changing
helicity allow easy determination of the compensation point, and
this is indicated by the arrows in each figure. As soon as the
$B_{z}$ field is swept away from the compensation point, nuclear
polarization may begin to occur.  Moving away from the compensation
point, the magnitude of generated nuclear field $B_N$ increases.
However, the magnitude of the $B_N$ field with respect to the
fluctuation field $B_f$ and the orientation of $B_N$ has a complex
effect on the resultant electron spin orientation, which we attempt
to clarify in the next section.

\subsection{Tuning the angle of the Overhauser field with milli-Tesla external fields}
\begin{figure}
\centering
\includegraphics[width=\columnwidth]{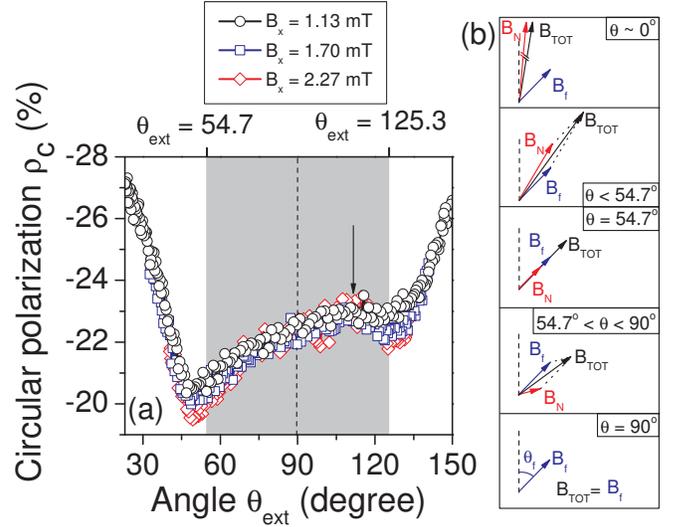}
\caption{(a) Data from Fig.\ref{figure8} (b) replotted as a function
of angle of applied field, $\theta_{ext} = \arctan(B_x/B_z)$.  Arrow
shows approximate position of the Knight field compensation point.
(b) series of schematics showing the magnitude and direction of the
relevant nuclear fields for important regions of the applied field
angle.  Blue arrow: fluctuation field, $\mathbf{B}_f$; Red arrow:
optically generated Overhauser field $\mathbf{B}_N$; Black arrow:
resultant total nuclear field
$\mathbf{B}_{tot}=\mathbf{B}_{f}+\mathbf{B}_{N}$} \label{figure9}
\end{figure}

The experiment shown in Fig.~\ref{figure8} involves choosing an
external $B_x$ field, which is kept constant, and sweeping another
$B_z$ field, from negative to positive values, through zero.  We
therefore effectively sweep the angle of the external field from
along the $x$-axis to along the $z$-axis, as explained previously.
To further elucidate our data we have taken the {\it same} data
shown in Fig.~\ref{figure8} (b) and {\it replotted} it, not as a
function of $B_z$, but as a function of the angle of the total
external field, as shown in Fig.~\ref{figure9}.  This angle is given
by $\theta_{ext} = \arctan(B_x/B_z)$.  Note that $\theta_{ext} =
90\degree$ and $0\degree$ corresponds to the external field along
the $x$ and $z$ axes, respectively.
Upon replotting the data, it is evident that $\rho_c$ is dependent
on $\theta_{ext}$ and not on the absolute magnitudes of $B_x$ or
$B_z$.  The curves for $B_x =$ 1.13, 1.70 and 2.27~mT obviously
coincide, and show the same asymmetry as well as the W-shaped
feature.  The value of $\rho_c$ at the Knight field compensation
point is $\sim$~-23~\% (indicated by the arrows).  Moving away from
this point a {\it reduction} in $\rho_c$ is observed, until a
turning point is reached, and $\rho_c$ then increases sharply.  We
now explain this behavior qualitatively.

In this experiment, the magnitude of the Overhauser field $B_N$
generated, is small, unless the applied $B_z$ field is very large.
Therefore we are always in the regime where the Overhauser field
generated is of the same order of magnitude as the fluctuation field
($B_N \sim B_f$), and the two are in direct competition.  At the
Knight field compensation point, the magnitude of $B_N$ is at its
lowest, and the electron sees a pure $B_f$ field.  The electron
precesses around this fluctuation field, at $\theta_f = 54.7\degree$
(first panel from bottom in Fig 9(b)).

Now let us consider what happens as we move away from $\theta_{ext}
= 90\degree$ towards the value 54.7$\degree$.  We see that the
polarization decreases, indicating that electron spin projection
onto $z$ decreases.  This appears counterintuitive.  If the
Overhauser field angle $\theta(\mathbf{B}_N)$ of the {\it polarized}
nuclear field is becoming closer to the $z$-axis, the electron
polarization should increase.  The second panel reveals why $S_z$
decreases in this region.  The nuclear field generated for
$\theta_{ext} > 54.7\degree$ is much smaller in magnitude than
$B_f$, but as $\theta_{ext}$ is decreased, the {\it magnitude} of
$\mathbf{B}_N$ increases (due to the fact that $S_z$ is larger) and
begins to compete more strongly with $\mathbf{B}_f$.  It is clear
from the second panel in Fig. 9(b) that for $\theta_{ext} >
54.7\degree$, the total field $\mathbf{B}_{tot} = \mathbf{B}_f +
\mathbf{B}_N$ is at a larger angle than $B_f$.  This means that the
stronger the Overhauser field $\mathbf{B}_N$ generated, the more the
electron depolarizes.

At $\theta = 54.7 \degree$ a turning point is reached.  At this
point, $\mathbf{B}_f$ and $\mathbf{B}_N$ are at the same angle (see
the third panel of Fig. 9(b)), and increasing $B_N$ is no longer
detrimental.  Upon increasing $\theta_{ext}$ further, any increase
in $\mathbf{B}_N$ leads to a total field $\mathbf{B}_{tot}$ which is
always at an angle smaller than $54.7\degree$.  This has a positive
feedback effect: the electron spin is preserved, and therefore may
polarize more nuclei.  As $\theta_{ext}$ is decreased further,
$B_N$, and hence $S_z$ increase quickly, as depicted in the final
panel at the top.

We have described the behavior shown in Fig.~\ref{figure8} in a
quantitative way only.  A qualitative description would require
detailed knowledge of $\mathbf{B}_N$ as a function of angle, a value
which is likely to be non-linear, and is beyond the scope of this
paper.  However, it shows clear evidence that these small external
fields may be used to accurately fine-tune the angle of the
Overhauser field generated.  With improvements in nuclear pumping
rate one may be able to control this angle over even wider ranges.
The Overhauser field may therefore replace a strong external field
used to manipulate electron spins, ranging from the Voigt to the
Faraday geometry.

\subsection{Evaluation of Knight field and nuclear field}
\label{concl} Finally, in order to evaluate the Knight field we
determine in Fig.~\ref{figure8} the magnetic field at which the
Knight field was compensated. This compensation point has a position
of
\begin{equation}
|\tilde {\mathbf{B}}_e| = 0.5 \pm 0.1 \;\text{mT}.
\end{equation}
This agrees well with the value of 0.6 mT which was measured in single QD experiments \cite{lai2006}.

From Eq.~(\ref{knight1}) in Section IV.D, it was discussed that each
$i^{th}$ nucleus in the QD has an unique Knight field, $b^i_e$.  In
this experiment, a weighted average Knight field is measured: as
discussed in Section IV.D, one may approximate this weighted average
to be $\tilde {B}_e \sim B_{e,max}/2$ \cite{foot1}.  The weighted
average will clearly depend on the details of the electron
confinement within the QD, which goes beyond the scope of this
paper, but is not thought to deviate much from this approximation.
We recall also that the measured Knight field will be reduced
compared to the maximum obtainable Knight field, due to the fact
that the average electron spin projection $S_z$ is reduced.
Similarly to Eq.~(\ref{knight1})
\begin{equation}
\mathbf{\tilde B}_e  = -{\tilde b}_e \frac{\langle \mathbf{S}\rangle}{S}. \label{knightave}
\end{equation}

Let us now consider the compensation points $\tilde {\mathbf{B}}_e =
-\mathbf{B}_z$ carefully again (arrows in Fig 8).  At this point,
the external field and internal Knight fields cancel, and generation
of nuclear polarization is suppressed.  The only field from the
nuclei is now the fluctuation field $B_f$.  As discussed before, the
electron precesses about this field at $\theta = 54.7\degree$, and
thus, from Eq.~(\ref{onethird}), $S_z = S_0/3$.  From
Eq.~(\ref{knightave}), $\tilde {B}_e = \tilde{b}_e S_z/S_0 $, it
follows that:
\begin{equation}
\tilde {b}_e \sim 1.5 \pm 0.3~\text{mT}. \label{knight_max}
\end{equation}
The value $\tilde {b}_e$ gives the maximum Knight field onto the system if no depolarization of the electron spin occurs.

\begin{figure}
\centering
\includegraphics[width=\columnwidth]{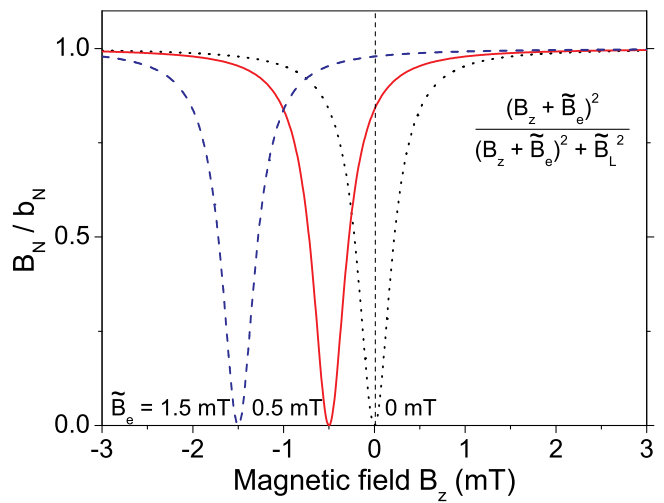}
\caption{$(B_z + \tilde {B}_e)^2/((B_z + \tilde {B}_e)^2+\tilde{B}_L^2)$ which is a measure for the competition between $\tilde{B}_L$ and $B_e$ concerning the ability of an Overhauser field to be generated.  Values $\tilde {B}_e =0$, 0.5 and 1.5~mT are shown.  The most important point to consider is the value at exactly $B_z = 0$ (vertical dashed line), where no external field supports nuclear polarization.}
\label{figure10}
\end{figure}
The information we have gained about the magnitude of the Knight
field and the dipole-dipole field now enable us to estimate the
magnitude of the maximum achievable nuclear magnetic field using
Eq.~(\ref{knight_simple}).  In the calculations a leakage factor $f$
accounting for phenomenological losses of nuclear spin polarization
not explicitly discussed here, was set to one so that the results
should be understood as the maximum polarization which may be
generated with the measured values $\tilde {B}_{e} = 0.5$ mT and
$\tilde{B}_{L} = 0.22$ mT.  Simply taking these values, one obtains
a value for the estimated maximum field of:
\begin{eqnarray}
\frac{B_N}{b_N} & \approx &  \frac{\tilde{B}_{e}^2}{\tilde{B}_{e}^2 + \tilde{B}_L^2} = 0.84.\label{knight_lower}
\end{eqnarray}
Thus we observe that even a moderate Knight field effectively
dominates over the dipole-dipole field, and up to $84 \%$ nuclear
polarization may be obtained in the absence of any other leakage.

One may calculate $b_N$ in the case that $100\%$ of the nuclear
spins are polarized.  For In$_{0.5}$Ga$_{0.5}$As QDs with electron
g-factor $g_e = 0.5$ \cite{greilich2006} it was estimated that
\cite{gueron1964,paget1977,oulton2007}
\begin{eqnarray}
\label{bni}
b_{N,In_{0.5}} & \approx & -4.3 \;T, \nonumber \\
b_{N,Ga_{0.5}} & \approx & -1.26 \;T, \nonumber\\
b_{N,As} & \approx & -2.76 \;T.
\end{eqnarray}
With these values we obtain a maximum nuclear field of $b_{N,max} =
\sum_{i} b_{N,i} = -8.3$ T.  This value is, in fact, exact for the
alloy composition given and is independent of electron localization.
From Eq.~(\ref{knight_lower}) therefore, we might expect to see an
Overhauser field of $B_N = 6.9$~T.

However, one should consider more carefully the value taken for the
Knight field.  In a system where little nuclear polarization has yet
built up, the electron precesses around the $B_f$ field and, as we
have shown, the residual electron spin polarization along $z$ gives
rise to a Knight field of $\sim 0.5$~mT.  It was in this regime that
the Knight field was directly measured here.  However, should a
moderate Overhauser field build up in the $z$-direction that
dominates over the fluctuating field, the electron does not precess
about an oblique field, and both the electron spin projection and
the Knight field reach their maximum values.  We have already shown
this maximum value to be $\tilde {b}_e \sim 1.5$~mT in this system.
If we make the assumption that at $B_{ext} = 0$ at least a moderate
Overhauser field builds up in many of our QDs, we may use the
maximum value of the Knight field ($B_e = b_e$)calculated from
Eq.~(\ref{knight_max}) to be
\begin{eqnarray}
\label{BN1}
\frac{B_N}{b_N} & \approx &  \frac{\tilde {b}_{e}^2}{\tilde {b}_{e}^2 + \tilde{B}_L^2} =0.98.
\end{eqnarray}
Thus, we observe that as soon as a significant Overhauser field
builds up in the QDs, the Knight field magnitude is at a maximum,
and one should theoretically be able to obtain almost 100~\% nuclear
spin alignment (and an Overhauser field of $B_N = 8.1$~T).

Note that the maximum projection of the Knight field onto the nuclei
is not necessarily reflected in the measured $\rho_c$ value.  The
$\rho_c$ value is also governed by the probability of electron
spin-flip from the wetting layer to the ground state during
relaxation.  Thus, the nuclear spins in a particular QD may be
prepared with a strong alignment in the $z$ direction.  An electron
in the ground state therefore will not precess and lose its spin
projection onto the $z$-axis.  This is regardless of whether it has
spin up or down.  Thus while the sign of the Knight field will
change if the electron is spin-flipped, the magnitude stays the
same.

By way of illustration, Fig.~\ref{figure10}. demonstrates the effect
that different values of the Knight field have on the maximum
obtainable nuclear polarization.  In this figure, the function
\begin{eqnarray}
\label{fig9eq}
\frac{B_N}{b_N} & \approx &  \frac{(\mathbf{B}_z + \mathbf{\tilde {B}}_{e})^2}{(\mathbf{B}_z + \mathbf{\tilde {B}}_{e})^2 + \tilde{B}_L^2}.
\end{eqnarray}
is plotted as a function of $B_z$ for Knight field values of $\tilde
{B}_e = 0, 0.5, 1.5$~mT, where $\tilde{B}_L$ is taken to be $0.22
$~mT from Fig.~\ref{figure5}.  The vertical dashed line indicates
the value $B_z = 0$.  It is clear here that for no Knight field, no
nuclear polarization will occur according to this simple model, but
for values of $\tilde{B}_e$ determined in this work, the nuclear
polarization should reach large values.

In our previous work on the same sample \cite{oulton2007} strong
evidence was found for high Overhauser fields, that allow the
formation of a nuclear spin polaron state.  The measurements here
demonstrate that almost 100~\% alignment should indeed be possible
in these QDs.  Note that we make the assumption that without
dipole-dipole depolarization, 100~\% alignment would be achieved.
Clearly, the maximum nuclear polarization achievable is dependent on
several factors, of which the Knight field magnitude is just one.

\section{Summary and conclusions}
We have demonstrated that the effect of negative circular
polarization may be used both to polarize the spins of the resident
electrons in n-doped QDs and to optically orient the nuclear spins
in the QDs via spin transfer from the spin oriented electrons.
Furthermore, the spin polarization of the resident electrons may be
read out by measuring the circular polarization of the
photoluminescence upon circularly polarized non-resonant excitation
of the QDs.

The electron spin polarization at the same time serves as a
sensitive detector for the state of the nuclear spin system.
Milli-Tesla external magnetic fields may be used to manipulate the
nuclear spins which in turn amplify the external field by orders of
magnitude making it possible to detect their action via the electron
polarization. Exploiting this co-dependence of electron and nuclear
spins, we studied Hanle curves for excitation modulated between
$\sigma^+$ and $\sigma^-$ helicity with different modulation
periods.  We were able to show that it takes tens of milliseconds to
maximally polarize the nuclear spin system in the QDs using our
polarization method. It became obvious that one has to examine
thoroughly whether Hanle measurements in a specific case may be used
to determine spin dephasing times of the electron. Even when nuclear
polarization is suppressed by modulated excitation, the random
frozen fluctuation nuclear field is still present, dominating the
dynamics of the electron spins.  Thus one may conclude that
determining the electron spin lifetime using the Hanle effect for
magnetic fields less than a few tens of milli-Tesla is incorrect due
to screening either from the fluctuation field or the dynamic
polarization: an alternative method must be used in QDs.

The Hanle measurements, however, allowed us to determine the
dipole-dipole field $\tilde{B}_{L} \approx 0.22$ mT. Furthermore,
the magnetic field dependence of the PL polarization in a
combination of Faraday and Voigt geometries could be used to obtain
an accurate determination of the magnitude of the Knight field
$\tilde {b}_{e} \approx 1.5$ mT.  It was also demonstrated that one
may fine-tune the angle of $B_N$ to the $z$-axis.

After having determined the values of the dipole-dipole and Knight
fields for this one system, the maximum nuclear field achievable may
be calculated.  It was found that, neglecting losses, the nuclear
field at zero externally applied field may be as high as $\approx
6.9$~T, which is achievable due to the stabilizing influence of the
Knight field. It was also calculated that the nuclear polarization
could reach $>$~98~\% for a fully polarized electron spin, as the
maximum Knight field value of $\approx$ 1.5 mT dominates over the
dipole-dipole field.  This nuclear field, reaching up to 8.1~T
provides an explanation for the observation of polaron formation at
$T = 2$~K, as theoretically predicted \cite{merkulov1998} and for
which some experimental evidence already has been provided
\cite{oulton2006,oulton2007}.

The fact that we use a QD ensemble for our studies may be considered
a disadvantage because of the ensemble inhomogeneities.  However,
the variation of the parameters of the electron-nuclear spin system
we measure is not necessarily primarily due to the distribution in
the ensemble but vary for a single QD due to the inhomogeneity over
the dot volume. For a single QD, the nuclear configuration may be
very different each time it is probed, and because the nuclear spins
may remain frozen for microseconds to seconds, one must integrate
over very long times to ensure true averaging effects. The fact that
we do see collective effects in our sample is a proof that the
ensemble broadening is relatively small concerning the parameters of
the electron-nuclear spin system.     On the contrary, it is a
remarkable finding that the ensemble reacts collectively yielding
the pronounced features we have observed.

To summarize, it is clear that the electron-nuclear system may be
manipulated with just a few milli-Tesla, in stark contrast to
conventional semiconductor systems.  The dynamics of this complex
system is only beginning to be understood, but clearly holds the key to achieving long electron spin qubit coherence times for use in applications such as quantum information processing, whilst the Knight field plays a crucial role in novel schemes for the use of QD nuclear spins as a quantum memory\cite{kurucz2009}.

\section*{Acknowledgments}
This work was supported by the Deutsche Forschungsgemeinschaft
(grant BA 1549/12-1) and the BMBF research program ``Nanoquit''.
S.~Yu.~Verbin and R.~V.~Cherbunin acknowledge support of the Russian
Ministry of Science and Education (grant RNP.2.1.1.362) and the
Russian Foundation for Basic Research, R.~Oulton thanks the
Alexander von Humboldt Foundation.



\begin{references}

\bibitem[$\star$]{thomasmail} e-mail: \texttt{thomas.auer@e2.physik.uni-dortmund.de}

\bibitem[$\dag$]{ruthmail} Present address: Centre for Quantum Photonics, University of Bristol, Bristol BS8 1UB, United Kingdom. \\e-mail: \texttt{ruth.oulton@bristol.ac.uk}

\bibitem[$\ddag$]{dimaioffe} Also in: A.~F.~Ioffe Physico-Technical Institute, Russian Academy of Sciences, 194021 St.~Petersburg, Russia

\bibitem{lampel1968} G.~Lampel, Phys.~Rev.~Lett.~{\bf 20}, 491 (1968).

\bibitem{overhauser1953} A.~W.~Overhauser,
Phys.~Rev.~{\bf 92}, 411 (1953).

\bibitem{ekimovsafarov1972} A.~I.~Ekimov and V.~I.~Safarov, ZhETF Pis.~Red.~{\bf 15}, 453 (1972) [JETP Lett.~{\bf 15}, 319 (1972)].

\bibitem{berkovits1973} V.~L.~Berkovits, A.~I.~Ekimov, and V.~I.~Safarov, Zh.~Eksp.~Teor.~Fiz.~{\bf 65}, 346 (1973) [Sov.~Phys.~JETP {\bf 38}, 169 (1974)].

\bibitem{dyakonov1974} M.~I.~D'yakonov, V.~I.~Perel', V.~L.~Berkovits, and V.~I.~Safarov, Zh.~Eksp.~Teor.~Fiz.~{\bf 67}, 1912 (1974) [Sov.~Phys.~JETP {\bf 40}, 950 (1975)].

\bibitem{dyakonovperel1974} M.~I.~D'yakonov and V.~I.~Perel', Zh.~Eksp.~Teor.~Fiz.~{\bf 65}, 362 (1973) [Sov.~Phys.~JETP {\bf 38}, 177 (1974)].

\bibitem{fleisher1975} V.~A.~Novikov and V.~G.~Fleisher, Pis'ma Zh.~Tekh.~Fiz.~{\bf 20}, 935 (1975).

\bibitem{OptO} F.~Meier and B.~P.~Zakharchenya (editors), {\it Optical Orientation}, Modern Problems in Condensed Matter Sciences, Vol.~8 (North-Holland, Amsterdam, 1984).

\bibitem{merkulov2009} I.~A.~Merkulov, G.~Alvarez, D.~R.~Yakovlev and T.~C.~Schulthess, arXiv:0907.2661 (2009). 

\bibitem{paget1977}
D.~Paget, G.~Lampel, B.~Sapoval, and V.~I.~Safarov, Phys.~Rev.~B {\bf 15}, 5780 (1977).

\bibitem{ekimov1972} A.~I.~Ekimov and V.~I.~Safarov, ZhETF Pis.~Red.~{\bf 15}, 257 (1972) [JETP Lett.~{\bf 15}, 179 (1972)].

\bibitem{berkovits1978} V.~L.~Berkovits, C.~Hermann, G.~Lampel, A.~Nakamura, and V.~I.~Safarov, Phys.~Rev.~B {\bf 18}, 1767 (1978).

\bibitem{brown1996} S.~W.~Brown, T.~A.~Kennedy, D.~Gammon, and E.~S.~Snow,
Phys.~Rev.~B {\bf 54}, R17339 (1996).

\bibitem{gammon2001} D.~Gammon, Al.~L.~Efros, T.A.~Kennedy, M.~Rosen, D.S.~Katzer, and D.~Park,
Phys.~Rev.~Lett.~{\bf 86}, 5176 (2001).

\bibitem{koppens2005} F.H.L.~Koppens, J.A.~Folk, J.M.~Elzerman, R.~Hanson, L.H.~Willems van Beveren, I.T.~Vink, H.P.~Tranitz, W.~Wegscheider, L.P.~Kouwenhoven, and L.M.K.~Vandersypen, Science {\bf 309}, 1346 (2005)

\bibitem{akimov2006} I.~A.~Akimov, D.~H.~Feng, and F.~Henneberger,
Phys.~Rev.~Lett.~{\bf 97}, 056602 (2006).

\bibitem{eble2006} B.~Eble, O.~Krebs, A.~Lemaître, K.~Kowalik, A.~Kudelski, P.~Voisin, B.~Urbaszek, X.~Marie, and T.~Amand,
Phys.~Rev.~B {\bf 74}, 081306 (2006).

\bibitem{braun2006} P.-F.~Braun, B.~Urbaszek, T.~Amand, X.~Marie, O.~Krebs, B.~Eble, A.~Lemaitre, and P.~Voisin,
Phys.~Rev.~B {\bf 74}, 245306 (2006).

\bibitem{tartakovskii2007} A.~I.~Tartakovskii, T.~Wright, A.~Russell, V.~I.~Fal'ko, A.~B.~Van'kov, J.~Skiba-Szymanska, I.~Drouzas,1 R.~S.~Kolodka, M.~S.~Skolnick, P.~W.~Fry, A.~Tahraoui, H.-Y.~Liu, and M.~Hopkinson,
Phys.~Rev.~Lett.~{\bf 98}, 026806 (2007).

\bibitem{feng2007}
D.~H.~Feng, I.~A.~Akimov, and F.~Henneberger,
Phys.~Rev.~Lett.~{\bf 99}, 036604 (2007).

\bibitem{maletinsky2007} P.~Maletinsky, A.~Badolato, and A.~Imamoglu,
Phys.~Rev.~Lett.~{\bf 99}, 056804 (2007).

\bibitem{oulton2007} R.~Oulton, A.~Greilich, S.~Yu.~Verbin, R.~V.~Cherbunin, T.~Auer, D.~R.~Yakovlev, M.~Bayer, I.~A.~Merkulov, V.~Stavarache, D.~Reuter, and A.~D.~Wieck , Phys.~Rev.~Lett.~{\bf 98}, 107401 (2007).

\bibitem{greilich2006}  A.~Greilich, D.~R.~Yakovlev, A.~Shabaev, Al.~L.~Efros, I.A.~Yugova, R.~Oulton, V.~Stavarache, D.~Reuter, A.~Wieck, and M.~Bayer, Science {\bf 313}, 341 (2006).

\bibitem{epstein2001} R.~J.~Epstein, D.~T.~Fuchs, W.~V.~Schoenfeld, P.~M.~Petroff, and D.~D.~Awschalom,
Appl.~Phys.~Lett.~{\bf 78}, 733 (2001).

\bibitem{bracker2005} A.~S.~Bracker, E.~A.~Stinaff, D.~Gammon, M.~E.~Ware, J.~G.~Tischler, A.~Shabaev, Al.~L.~Efros, D.~Park, D.~Gershoni, V.~L.~Korenev, and I.~A.~Merkulov, Phys.~Rev.~Lett.~{\bf 94}, 047402 (2005).

\bibitem{lai2006} C.~W.~Lai, P.~Maletinsky, A.~Badolato, and A.~Imamoglu,
Phys.~Rev.~Lett.~{\bf 96}, 167403 (2006).

\bibitem{fafard1999} S.~Fafard and C.~Allen, Appl.~Phys.~Lett.~{\bf 75}, 2374 (1999).

\bibitem{cortez2002} S.~Cortez, O.~Krebs, S.~Laurent, M.~Senes, X.~Marie, P.~Voisin, R.~Ferreira, G.~Bastard, J-M.~Gérard, and T.~Amand,
Phys.~Rev.~Lett.~{\bf 89}, 207401 (2002).

\bibitem{laurent2006} S.~Laurent, M.~Senes, O.~Krebs, V.~K.~Kalevich, B.~Urbaszek, X.~Marie, T.~Amand, and P.~Voisin,
Phys.~Rev.~B {\bf 73}, 235302 (2006).

\bibitem{verbin2005} M.~Ikezawa, B.~Pal, Y.~Masumoto, I.~V.~Ignatiev, S.~Yu.~Verbin, and I.~Ya.~Gerlovin,
Phys.~Rev.~B {\bf 72}, 153302 (2005).

\bibitem{ware2005} M.~E.~Ware, E.~A.~Stinaff, D.~Gammon, M.~F.~Doty, A.~S.~Bracker, D.~Gershoni, V.~L.~Korenev, S.~C.~Badescu, Y.~Lyanda-Geller, and T.~L.~Reinecke, Phys.~Rev.~Lett.~{\bf 95}, 177403 (2005).

\bibitem{masumoto2006} Y.~Masumoto, S.~Oguchi, B.~Pal, and M.~Ikezawa,
Phys.~Rev.~B {\bf 74}, 205332 (2006).

\bibitem{kalevich2005}
V.~K.~Kalevich, I.~A.~Merkulov, A.~Yu.~Shiryaev, K.~V.~Kavokin, M.~Ikezawa, T.~Okuno, P.~N.~Brunkov, A.~E.~Zhukov, V.~M.~Ustinov, and Y.~Masumoto,
Phys.~Rev.~B {\bf 72}, 045325 (2005).

\bibitem{akimov2005} I.~A.~Akimov, K.~V.~Kavokin, A.~Hundt, and F.~Henneberger,
Phys.~Rev.~B {\bf 71}, 075326 (2005).

\bibitem{khaetskii2000} A.~V.~Khaetskii and Y.~V.~Nazarov
Phys.~Rev.~B {\bf 61}, 12639 (2000).

\bibitem{merkulov2002} I.~A.~Merkulov, Al.~L.~Efros, and M.~Rosen,
Phys.~Rev.~B {\bf 65}, 205309 (2002).

\bibitem{imamoglu2003}  A.~Imamoglu, E.~Knill, L.~Tian, and P.~Zoller,
Phys.~Rev.~Lett.~{\bf 91}, 017402 (2003).

\bibitem{stepanenko2006} D.~Stepanenko, G.~Burkard, G.~Giedke, and A.~Imamoglu,
Phys.~Rev.~Lett.~{\bf 96}, 136401 (2006).

\bibitem{deng2006} C.~Deng, and X.~Hu, Phys.~Rev.~B {\bf 73}, 241303(R) (2006).

\bibitem{dzhioev1999} R.~I.~Dzhioev, B.~P.~Zakharchenya, V.~L.~Korenev, and M.~V.~Lazarev,
Phys.~Solid State {\bf 41}, 2014 (1999).

\bibitem{braun2005} P.-F.~Braun, X.~Marie, L.~Lombez, B.~Urbaszek, T.~Amand, P.~Renucci, V.~K.~Kalevich, K.~V.~Kavokin, O.~Krebs, P.~Voisin, and Y.~Masumoto,
Phys.~Rev.~Lett.~{\bf 94}, 116601 (2005).

\bibitem{greilich2007} A.~Greilich, A.~Shabaev, D.R.~Yakovlev, Al.~L.~Efros, I.A.~Yugova, D.~Reuter, A.D.~Wieck, and M.~Bayer,
Science {\bf 317}, 1896 (2007).

\bibitem{abragam1983} A.~Abragam, {\it The Principles of Nuclear Magnetism}, (Clarendon Press, Oxford, 1983)

\bibitem{ignatiev2007} R.~V.~Cherbunin, S.~Yu. Verbin, T.~Auer, D.~R.~Yakovlev, D.~Reuter,  A.~D.~Wieck, I. Ya. Gerlovin, I.~V.~Ignatiev, D. V. Vishnevsky and M.~Bayer, Phys.~Rev.~{\bf 80}, 035326 (2009).

\bibitem{oulton2006} R.~Oulton, S.~Yu.~Verbin, T.~Auer, R.~V.~Cherbunin, A.~Greilich, D.~R.~Yakovlev, M.~Bayer, D.~Reuter, and A.~Wieck,
Phys.~Stat.~Sol.~(b) {\bf 243}, 3922 (2006).

\bibitem{knight1949} W.~D.~Knight, Phys.~Rev.~{\bf 76}, 1259 (1949).

\bibitem{ThThesis} T.~Auer, PhD.~Thesis: ``The electron-nuclear spin system in (In,Ga)As quantum dots'' (Sierke Verlag, Goettingen, 2008).

\bibitem{foot1}This weighted average reflects the fact that, while for a nucleus at a distance $r = a_0 / 2$ the Knight field value is $e^{-1}$, the contribution this nucleus makes to the Overhauser field is $e^{-1}$ of that of a nucleus at the centre.  Thus the average Knight field value measured is fairly narrow and weighted heavily towards the nuclei in the QD center.

\bibitem{hanle1924} W.~Hanle, Z.~Phys.~{\bf 30}, 93 (1924).

\bibitem{gueron1964}  M.~Gueron, Phys.~Rev.~{\bf 135}, A200 (1964).

\bibitem{merkulov1998} I.~A.~Merkulov, Phys.~Solid State {\bf 40}, 930 (1998).

\bibitem{comment1} The dominant mechanism for the evolution of the nuclear spins over this timescale is still under debate: both direct precession of the nuclear spins in the electronic field and electron mediated nuclear-nuclear spin-flip processes are thought to play a role. In any case, it is in fact the electron itself that causes nuclear spin evolution \cite{maletinsky2007}: without an electron present only the dipole-dipole interaction is important.

\bibitem{comment2} Note that the maximum and minimum values of $\rho_c$ vary slightly between Figs.~\ref{figure5} and \ref{figure6}, with the maximum varying from $-26 to -30$~\%.  This is because the measurements are taken on different measurement runs.  As we have demonstrated in reference \cite{oulton2007}, slight temperature variations in the bath cryostat give rise to different values of nuclear polarization decay over very long times.  The behavior of the changes in polarization in this paper when a magnetic field is swept are not strongly dependent on temperature, however.  Additionally, slight changes in excitation wavelength result in a change in electron spin memory preserved during energy relaxation from the wetting layer.  Again, this slight variation does not affect the electron-nuclear spin dynamics here.

\bibitem{kurucz2009}  Z.~Kurucz, M.~W.~Sorensen, J.~M.~Taylor, M.~D.~Lukin, M.~Fleischhauer, Phys.~Rev.~Lett.~{\bf 103}, 010502 (2009).

\end{references}
\end{document}